\documentclass[aps,prb,twocolumn,superscriptaddress]{revtex4-2}
\usepackage{amsfonts, amssymb, amsmath,amsthm,graphicx}
\usepackage{subfigure}
\usepackage{multirow}
\usepackage{dcolumn}
\usepackage{bm}
\usepackage{color}
\usepackage[usenames,dvipsnames]{xcolor}

\renewcommand{\bf}[1]{\textnormal{\textbf{#1}}}
\newcommand{\BZ}{\textnormal{\text{BZ}}}

\newcommand{\ket}[1]{| #1 \rangle}
\newcommand{\bra}[1]{\langle #1|}

\usepackage[colorlinks=true,allcolors=blue]{hyperref}

\begin{document}

\title{Engineering geometrically flat Chern bands with Fubini-Study K\"ahler structure}

\author{Bruno Mera}
\affiliation{Instituto de Telecomunica\c{c}\~oes, 1049-001 Lisboa, Portugal}
\affiliation{Departmento de F\'{i}sica, Instituto Superior T\'ecnico, Universidade de Lisboa, Av. Rovisco Pais, 1049-001 Lisboa, Portugal}
\affiliation{Departmento de Matem\'{a}tica, Instituto Superior T\'ecnico, Universidade de Lisboa, Av. Rovisco Pais, 1049-001 Lisboa, Portugal}
\author{Tomoki Ozawa}
\affiliation{Advanced Institute for Materials Research (WPI-AIMR), Tohoku University, Sendai 980-8577, Japan}

\date{\today}

\newcommand{\tom}[1]{{\color{red} #1}}
\definecolor{bbblue}{rgb}{0.,0.24,0.51}
\newcommand{\blue}{\color{bbblue}}

\begin{abstract}
We describe a systematic method to construct models of Chern insulators whose Berry curvature and the quantum volume form coincide and are flat over the Brillouin zone; such models are known to be suitable for hosting fractional Chern insulators. The bands of Chern insulator models where the Berry curvature and the quantum volume form coincide, and are nowhere vanishing, are known to induce the structure of a K\"ahler manifold in momentum space, and thus we are naturally led to define {\it K\"ahler bands} to be Chern bands satisfying such properties. We show how to construct a geometrically flat K\"ahler band, with Chern number equal to minus the total number of bands in the system, using the idea of K\"ahler quantization and properties of Bergman kernel asymptotics. We show that, with our construction, the geometrical properties become flatter as the total number of bands in the system is increased; we also show the no-go theorem that it is not possible to construct geometrically perfectly flat K\"ahler bands with a finite number of bands. We give an explicit realization of this construction in terms of theta functions and numerically confirm how the constructed K\"ahler bands become geometrically flat as we increase the number of bands. We also show the effect of truncating hoppings at a finite length, which will generally result in deviation from a perfect K\"ahler band but does not seem to seriously affect the flatness of the geometrical properties.
\end{abstract}

\maketitle
\section{Introduction}
\label{sec: Introduction}
Chern insulators are prototypical lattice models of topological insulators~\cite{Hasan:2010,Qi:2011}. Without interparticle interactions, fermionic Chern insulators exhibit the integer quantum Hall effect, where the Hall conductivity is proportional to the first Chern number of the occupied bands. Including appropriate interactions, Chern insulators with certain filling factors are expected to become fractional Chern insulators, which exhibit the fractional quantum Hall effect~\cite{Tang:2011,Sun:2011,Neupert:2011,Renault:2011,Sheng:2011,Parameswaran:2013,Bergholtz:2013}. To theoretically study fractional Chern insulators on lattices, a natural guiding principle is to explore possible analogies from the existing studies of fractional quantum Hall states in the continuum. A challenge here is to find analogies between the Landau level physics in the continuum and the eigenstates of Chern insulators on lattices. In previous works, it has been noticed that when the geometrical properties of noninteracting Chern insulators fulfill certain conditions, we can draw good analogies between Chern insulators on lattices and the fractional quantum Hall states in the continuum~\cite{claa:lee:tho:qi:dev:15,lee:cla:tho:17}. In particular, it is desirable to have bands whose Berry curvature $F_{12}(\mathbf{k})$ and the quantum metric $g_{ij}(\mathbf{k})$ as a function of quasimomentum $\mathbf{k}$ in the Brillouin zone satisfy the equality $\sqrt{\det (g(\mathbf{k}))} = |F_{12}(\mathbf{k})/2|$. When this condition is met, the Bloch states of a Chern band become analogous to the lowest Landau levels in the continuum, from which we can expect stable construction of fractional Chern insulators. This condition is related to the possibility to choose Bloch states as holomorphic functions on momentum space. Earlier works have proposed to construct Chern insulators with such properties using variational states involving elliptic functions~\cite{claa:lee:tho:qi:dev:15,lee:cla:tho:17}. 

The condition of $\sqrt{\det (g(\mathbf{k}))} = |F_{12}(\mathbf{k})/2|$ together with $\det (g(\mathbf{k})) > 0$ is equivalent to momentum space being a K\"ahler manifold with the quantum metric as the K\"ahler metric and the Berry curvature (up to a constant) as the K\"ahler form~\cite{claa:lee:tho:qi:dev:15,lee:cla:tho:17,oza:mer:21:phys:published,mer:oza:21:math:published}; in this paper we define {\it K\"ahler bands} to be energy bands fulfilling these conditions. 
Here we propose a systematic method to construct a sequence of K\"ahler bands, labeled by the total number of bands in the system, with asymptotically flat geometry, that is, $\sqrt{\det (g(\mathbf{k}))} = |F_{12}(\mathbf{k})/2|$ is asymptotically constant over the Brillouin zone. We note that K\"ahler bands with flat energy dispersion have been discussed under the name of (near-)ideal flatbands in the recent works of Refs.~\cite{khalaf:vishwanath:20, wang:cano:millis:liu:yang:21}, in which the flatness of geometry is not imposed.

In our method, once we fix an auxiliary Hermitian holomorphic line bundle on the Brillouin zone, $L \to \mathrm{BZ}^2$, with its curvature satisfying certain conditions, we can construct K\"ahler bands from an orthogonal basis of the vector space of (global) holomorphic sections of the tensor-product line bundle $L^{\otimes p} \to \mathrm{BZ}^2$. The total number of bands of the constructed model equals the opposite of the Chern number of the constructed K\"{a}hler band which, in turn, is equal to $-p\mathcal{C}$, where $\mathcal{C}$, with $\mathcal{C}>0$, is the Chern number of $L\to\BZ^2$. The K\"ahler band approaches flat geometry in the limit of large $p$, which is guaranteed from a known mathematical theorem on the Bergman kernel asymptotics~\cite{zel:98}. Notably, in order to obtain flat K\"ahler bands, we do not need to determine any parameter in a variational manner. Our method, being exact in the infinite total number of bands limit, yields asymptotically geometrically flat bands with a high Chern number. This should be contrasted to typical studies of fractional Chern insulators which focus on bands with low Chern number, in particular, with Chern number equal to $1$ due to the analogy with lowest Landau level physics. More recent studies~\cite{and:neu:mol:21} are focusing on fractional Chern insulators in bands with higher Chern numbers as they have been recently realized experimentally~\cite{span:etal:18} and they can exhibit richer physical phenomena. We confirm the validity of our method by explicitly constructing a model using theta functions. Although our model generally contains long-range hoppings in real space to an arbitrary distance, we also numerically show that truncating the hopping at a reasonable length would not seriously affect the resulting geometry of the model. 

Below, in Sec.~\ref{sec:bandgeometry}, we first introduce the basic terminology of band geometry and describe how to introduce a complex structure in momentum space. In Sec.~\ref{sec: Background on fractional Chern insulators and band geometry}, we then review arguments on how Chern bands with certain geometrical properties are preferred to obtain fractional Chern insulators. In Sec.~\ref{sec: Quasi-Kahler two-band models and holomorphic maps from BZ2 to CP1}, we discuss the general expressions of K\"ahler (and quasi-K\"ahler) bands as holomorphic maps from the Brillouin zone to complex projective spaces. Section~\ref{sec: The Bergman kernel prescription} discusses a systematic construction of flat K\"ahler bands, which forms the central result of our paper. In Sec.~\ref{sec: Explicit construction}, we provide an explicit construction of K\"ahler bands based on our method, including numerical calculations. Finally, we present the conclusions in Sec.~\ref{sec: Conclusions}. Some detailed derivations are given in Appendix, including a proof that it is not possible to construct a perfectly flat K\"ahler band with a finite number of bands.

\section{Band geometry and complex structure}
\label{sec:bandgeometry}We first summarize basic terminology of band geometry and how it can give rise to a complex structure on the Brillouin zone.
The geometry of the Bloch states of a single isolated band is characterized by a map $P:\BZ^2 \to \mathbb{C}P^{n-1}$, where $\BZ^2$ is the two-dimensional Brillouin zone and $\mathbb{C}P^{n-1}$ is the $(n-1)$-dimensional complex projective space; physically $n$ corresponds to the total number of bands.
The map $P$ induces a metric on the Brillouin zone $ds^2=\sum_{i,j}g_{ij}(\bf{k})dk_idk_j$, where $\mathbf{k} \in \BZ^2$, called the \emph{quantum metric}, through the pullback of the standard Fubini-Study metric on $\mathbb{C}P^{n-1}$~\cite{provost:1980,page:1987}.
Similarly, from the pullback of the Fubini-Study symplectic form on $\mathbb{C}P^{n-1}$, we obtain a symplectic two-form $\omega = -iF/2$ on momentum space, whose only nonzero component is $\omega_{12} = -\omega_{21}$. The two-form $F$ is called the Berry curvature, whose integral gives the first Chern number of the band $\int_\mathrm{BZ^2}iF/2\pi\in\mathbb{Z}$.
Generally, an inequality between the determinant of the quantum metric, $\det (g) = g_{11}g_{22} - g_{12}^2$, and the Berry curvature holds: $\sqrt{\det (g(\bf{k}))} \ge |F_{12}(\bf{k})|/2$ as first noticed by Roy~\cite{roy:14}. Since $\sqrt{\det (g)} dk_1 \wedge dk_2$, if $\det(g)\neq 0$ everywhere, provides the natural volume form based on the quantum metric and the standard orientation of the Brillouin zone, we call this quantity {\it the quantum volume form} in this paper. We note, however, that although the quantum volume form is always nonnegative, it can generally be zero in certain points (regions) of the Brillouin zone, so it is not a volume form in the strict mathematical sense. From the quantum metric, it is sometimes possible to introduce a complex structure on the Brillouin zone which we proceed to describe.

First, we assume that the map $P$ is an immersion, in which case, as we show, a complex structure can always be defined on the entire Brillouin zone. Since $P$ is an immersion, the metric is everywhere nondegenerate on the Brillouin zone, namely $\det (g(\bf{k})) \neq 0$ for any $\bf{k} \in \BZ^2$. In this case the quantum metric is a Riemannian metric defined everywhere on the Brillouin zone. For a given point on the Brillouin zone, $\bf{k} \in \BZ^2$, it is known that there exists a system of coordinates $(u,v)$ in a neighborhood of $\mathbf{k}$ which satisfies $ds^2 = \sum_{ij}g_{ij}dk_i dk_j = \rho (du^2 + dv^2)$, where $\rho$ is a positive function. Such coordinates $(u,v)$ are called \emph{isothermal coordinates}, and by defining $z = u+iv$, where $(u,v)$ are taken consistent with the standard orientation on the Brillouin zone, we can introduce the complex coordinate $z$ on a neighborhood of $\mathbf{k}$. Such complex coordinates defined around different points on the Brillouin zone can be patched together to give a complex atlas of the Brillouin zone, giving the Brillouin zone the structure of a complex manifold of complex dimension one, namely a \emph{Riemann surface}. We note that for a given metric on the Brillouin zone and a choice of orientation, the choice of the local complex coordinate is unique up to local orientation preserving conformal transformations, i.e., up to local bi-holomorphisms---this ensures that the local coordinates patch together nicely in a holomorphic manner, and also it tells us that the complex structure on the Brillouin zone is uniquely determined by the quantum metric. Since the complex projective space $\mathbb{C}P^{n-1}$ is also a complex manifold, the map $P$ can now be considered as a map between complex manifolds.
As shown in Refs.~\cite{oza:mer:21:phys:published,mer:oza:21:math:published}, when $P$ is a holomorphic immersion, momentum space is a K\"ahler manifold with the quantum metric as the K\"ahler metric and the symplectic two form $\omega = -iF/2$ as the K\"ahler form. We thus define {\it K\"ahler bands} to be bands with $P$ being a holomorphic immersion, with respect to the complex structure on the Brillouin zone defined above. For K\"ahler bands, the equality $\sqrt{\det (g(\bf{k}))} = |F_{12}(\bf{k})|/2$ holds~\cite{lee:cla:tho:17,oza:mer:21:phys:published,mer:oza:21:math:published}.
A primary objective of this paper is to find {\it flat} K\"ahler bands, which are K\"ahler bands whose geometrical properties, i.e., the quantum metric, the symplectic form (which equals the quantum volume form) and the complex structure, are flat, i.e., their components in the (periodic) coordinates $(k_1,k_2)$ are constant.

We stress that the word \emph{flat} in the present work refers to the situation where the geometric structures of interest are constant with respect to the $(k_1,k_2)$ coordinates~\footnote{In a coordinate free description, the periodic $(k_1,k_2)$ coordinates give rise to two independent globally defined vector fields over the Brillouin zone $X_1=\partial/\partial k_1$ and $X_2=\partial/\partial k_2$. Then, what we mean by \emph{flatness} of a Kähler structure over the Brillouin zone is that the Lie derivatives, with respect to these vector fields, of the symplectic form $\omega$, the complex structure $j$ and the metric $g$, denoted by, respectively, $\mathcal{L}_{X_i}\omega$, $\mathcal{L}_{X_i}j$ and $\mathcal{L}_{X_i}g$, $i=1,2$, vanish identically.}, and does not refer to flatness of the energy dispersion unless explicitly stated. We also want to stress that geometric flatness does not mean vanishing Berry curvature, i.e., it does not mean that the Berry connection is a flat connection; rather the Berry curvature takes a constant nonzero value over the Brillouin zone.

Next, we consider the situation where the map $P$ is not an immersion, in which case existence of the complex structure on the entire Brillouin zone is not always guaranteed. When $P$ is not an immersion, there are certain points on the Brillouin zone where $\det (g(\bf{k})) = 0$. Let us denote the (closed) set of points on the Brillouin zone where $\det (g(\bf{k})) = 0$ by $S = \{ \bf{k} \in \BZ^2 |\det (g(\bf{k})) = 0 \}$. If $\BZ^2\backslash S$, namely the set of points where $\det (g(\bf{k})) \neq 0$, has more than one connected component, with each component necessarily separated by a loop, it is not possible to define a complex structure throughout the Brillouin zone.
We note that since the Berry curvature can change sign from one connected component to the other, the choice of local orientations in each connected component, based on the sign of the Berry curvature (more precisely using the pullback under $P$ of the Fubini-Study symplectic form as a volume form), cannot in general be made globally consistent over the whole Brillouin zone---something that would necessarily happen if the map was holomorphic or anti-holomorphic (a standard argument for this is presented Sec.~\ref{sec: Quasi-Kahler two-band models and holomorphic maps from BZ2 to CP1}). Over each connected component of $\BZ^2\backslash S$, we can find local complex coordinate systems consistent with the quantum metric by finding a local complex coordinate in sufficiently small open neighbourhoods of each point on $\BZ^2\backslash S$ as in the first case and patching them together. Whether or not the complex coordinates defined in this way on $\BZ^2\backslash S$ belong to some complex atlas on the entire Brillouin zone and, hence, come from a globally defined complex structure, depends on specific situations. If the map $P:\BZ^2 \to \mathbb{C}P^{n-1}$ is known to be holomorphic \emph{a priori} with respect to some complex structure $j$, then we know that we can extend the complex structure on $\BZ^2\backslash S$ as determined from the quantum metric to be defined on the entire $\BZ^2$ since the former will just be the restriction of $j$ to $\BZ^2\backslash S$.
We define {\it quasi-K\"ahler bands} to be bands where $P$ is not an immersion but we can define the complex structure on the entire Brillouin zone and thus $P$ is represented by a holomorphic function.
Also for the quasi-K\"ahler bands, the equality $\sqrt{\det (g(\bf{k}))} = |F_{12}(\bf{k})|/2$ holds throughout the Brillouin zone.
Note that since zeros of holomorphic functions are isolated, if $P$ is holomorphic (and non-constant) then $S$, which is determined by the simultaneous vanishing of the first derivatives of the local holomorphic functions determining $P$ locally---hence determined by zeros of (non-constant) holomorphic functions---, will consist of a collection of isolated points in the Brillouin zone and, due to compactness, this collection will necessarily be finite.

The quantum metric $g(\mathbf{k})$ is physically related to localization of states~\cite{Marzari:1997, Souza:2000, Ozawa:2019}, and it is experimentally observable. Recent experiments have reported measurements of the quantum metric in various setups of synthetic quantum matter~\cite{Asteria:2019,Gianfrate:2020}. The complex structure, which as we saw is determined from the quantum metric, in turn, is related to the anisotropy in localization of the insulating state where the entire band is occupied by fermions~\cite{Mera:2020}.

\section{Background on fractional Chern insulators and band geometry}
\label{sec: Background on fractional Chern insulators and band geometry}
Following Ref.~\cite{lee:cla:tho:17}, we now recall some notions on fractional Chern insulators and the role of band geometry on their stability.

Consider the Hamiltonian for an isotropic free electron gas in two-dimensions in the presence of an external uniform magnetic field:
\begin{align*}
H=\frac{1}{2m}\sum_i\left( \bf{p}_i -e\bf{A}_i\right)^2 +\sum_{i<j}V(\bf{r}_i-\bf{r}_j),
\end{align*}
where $m$ is the effective mass, $e\bf{A}=(1/2)(-y,x)$ is the electromagnetic gauge field in the symmetric gauge and we have set the magnetic length $\ell_{B}^2=1/eB=1$. This is the canonical model for the fractional quantum Hall effect. Assume filling fraction $\nu<1$ and assume that the cyclotron frequency, i.e., the Landau level gap, is much larger than the scale of interactions, and, hence, an effective description based on projection onto the lowest Landau level (LLL) subspace is valid. The single-body electron problem can be written as
\begin{align*}
H_0=\omega_c(b^\dagger b +1/2),
\end{align*}
where $\omega_c=e B/m$ is the cyclotron frequency, $b=\left(-i\partial_{\overline{z}}-iz/2\right)/\sqrt{2}=(-i/\sqrt{2})\nabla_{\partial/\partial\overline{z}}$ is the Cauchy-Riemann operator determined by the covariant derivative $\nabla = d-i(xdy-ydx)/2=d+(zd\overline{z}-\overline{z}dz)/4$, where $d$ is the exterior derivative, and the isotropic flat complex structure determined by the complex coordinate $z=x+iy$ in the plane. The LLL wave functions are then square-integrable holomorphic functions, with respect to an appropriate inner product~\cite{gir:jac:84}, which can also be understood as holomorphic sections of the electromagnetic gauge bundle $L=\mathbb{R}^2\times\mathbb{C}\to \mathbb{R}^2$ which is equipped with the connection $\nabla=d+A$, $A=-i\bf{A}^{\flat}=-i(xdy-ydx)/2$, where $\flat$ is the musical isomorphism sending vector fields to $1-$forms. There is another oscillator algebra associated with the single-particle problem, namely,
\begin{align*}
a=-i\left(\partial_z +z/2\right)/\sqrt{2},
\end{align*}
which commutes with the one determined by $b$. Because of this, if $\ket{\psi}$ is in the LLL so is $a^{\dagger}\ket{\psi}$. One can then build all the states in the LLL from the vacuum of the $a$'s within the LLL, which is the state that satisfies $a\ket{\Psi_0}=b\ket{\Psi_0}=0$. This state is most simply the Gaussian $\Psi_0(z)\sim e^{-|z|^2/4}$. Laughlin's many-body trial wave functions may now be elegantly expressed in terms of the $a$'s (one for each particle):
\begin{align*}
\Psi_{\text{Laughlin}}=\prod_{i<j}\left(a_i^{\dagger}-a_j^{\dagger}\right)^{1/\nu} \Psi_0^{\otimes N},
\end{align*}
where $N$ is the number of particles and $n=1/\nu$ is odd (to satisfy Fermi statistics). The choice of the guiding-center basis $\left(\left(a^{\dagger}\right)^{m}/\sqrt{m!}\right)\Psi_0\sim z^{m}e^{-|z|^2/4}$, eigenstates of $L_z$, makes perfect sense in the isotropic case. However, when the effective mass is anisotropic this choice is not appropriate. The anisotropy is related to a choice of a \emph{complex structure} in the plane, independent of the kinetic energy and the interacting potential (which is assumed to be isotropic), which enters as a variational degree of freedom determining the geometry of the ground state wave function.

For Chern insulators, one is forced to abandon isotropy in order to be able to have a fractional quantum Hall fluid on the lattice. While Bloch bands, labeled by $\bf{k}\in\BZ^2$, seem to have a different structure from the wave functions in the LLL, one can still define an analogous guiding-center basis on the lattice. The way to do this is to define the FQHE states in terms of a basis of eigenstates of a small anisotropic confining potential which is projected to the flat Chern band~\cite{claa:lee:tho:qi:dev:15}. Namely, one makes the observation that while the angular momentum does not directly translate to the lattice, there is an equivalent description of the associated eigenstates, $\Psi_m=\left(\left(a^{\dagger}\right)^{m}/\sqrt{m!}\right)\Psi_0\sim z^{m}e^{-|z|^2/4}$, as eigenstates of a parabolic confinement potential $V(\bf{r})=(\lambda/2)\bf{r}^2=(\lambda/2)\bar{z}z$ projected to the $LLL$,
\begin{align*}
\left(P_{LLL}V(\bf{r})P_{LLL}\right)\Psi_n=\frac{\lambda}{2}(n+1)\Psi_n,\ n\in\mathbb{N},
\end{align*}
where $\lambda$ is a small real constant -- the latter having a well-defined lattice version. Indeed, we can swiftly adapt this construction to fractional Chern insulators by taking an anisotropic confinement potential on the lattice:
\begin{align*}
V(\bf{r})=\frac{1}{2} \lambda\; \eta_{ij}x^ix^j,
\end{align*}
where $\eta=(\eta_{ij})$ is a unimodular Galilean metric which is to be interpreted as a variational degree of freedom and where we assume the Einstein summation convention for repeated indices. Since under the Zak transform~\cite{zak:68} the position operators get mapped to $x^j=i\partial/\partial k_j$, $j=1,2$, it follows that the projection onto a Chern band, described by a rank $1$ smooth projector $P(\bf{k})$, yields
\begin{align*}
\overline{V}(\bf{k}) &\equiv P(\mathbf{k})\circ V(\bf{r})\circ P(\bf{k})
\\
&=-\frac{1}{2} \lambda\; \eta_{ij}P(\bf{k})\circ \frac{\partial}{\partial k_i}\circ \frac{\partial}{\partial k_j}\circ P(\bf{k}),
\end{align*}
where $\circ$ means operator composition. Expanding the derivatives, one finds
\begin{align}
\overline{V}(\bf{k})=-\frac{\lambda}{2}\eta_{ij}\nabla_{\frac{\partial}{\partial k_i}}\nabla_{\frac{\partial}{\partial k_j}} +\frac{\lambda}{2} \eta_{ij} g_{ij}(\bf{k}),
\label{eq: projected confining potential}
\end{align}
where $\nabla=P\circ d\circ P$ and $g_{ij}(\bf{k})$ are, respectively, the Berry connection and the components of the quantum metric of the band under consideration. The metric $\eta$ identifies a flat complex structure on $\BZ^2$, described by a (multi-valued) complex coordinate $z=k_1+\tau k_2$, through the formula
\begin{align}
\eta^{ij}dk_idk_j=\eta^{11}|dk_1 +\tau dk_2|^2=\eta^{11}|dz|^2, \ \tau\in\mathcal{H},
\label{eq: eta and tau}
\end{align}
where $\eta^{ij}$ are the matrix elements of the inverse matrix $\eta^{-1}$.
Here $\mathcal{H}$ is the upper half of the complex plane, and thus $\mathrm{Im}(\tau) > 0$.
We can then write 
\begin{align*}
\eta_{ij}\nabla_{i}\nabla_{j}&=2\eta_{11}\left(\nabla_{z}\nabla_{\overline{z}}+\nabla_{\overline{z}}\nabla_{z}\right)\\
&=2\eta_{11}\left(2\nabla_{z}\nabla_{\overline{z}} -[\nabla_{z},\nabla_{\overline{z}}]\right)\\
&=4\eta_{11}\nabla_{z}\nabla_{\overline{z}} -2\eta_{11} F_{z\overline{z}},
\end{align*}
and, equivalently,
\begin{align*}
\eta_{ij}\nabla_{i}\nabla_{j}&=2\eta_{11}\left(\nabla_{z}\nabla_{\overline{z}}+\nabla_{\overline{z}}\nabla_{z}\right)\\
&=2\eta_{11}\left(2\nabla_{\overline{z}}\nabla_{z} -[\nabla_{\overline{z}},\nabla_{z}]\right)\\
&=4\eta_{11}\nabla_{\overline{z}}\nabla_{z} +2\eta_{11} F_{z\overline{z}},
\end{align*}
where $F=F_{z\overline{z}}dz\wedge d\bar{z}=(\overline{\tau}-\tau)F_{z\overline{z}}dk_1\wedge dk_2=F_{12}dk_1\wedge dk_2$ is the Berry curvature, so that we have the following two equivalent expressions for $\overline{V}(\bf{k})$
\begin{align}
\overline{V}(\bf{k})&=-2\lambda\eta_{11}\nabla_{z}\nabla_{\overline{z}} +\lambda\left(\frac{i\eta_{11}}{2\tau_2}F_{12} +\frac{1}{2}\eta_{ij}g_{ij}\right)
\label{eq: hol V}
\end{align}
and
\begin{align}
\overline{V}(\bf{k})&&=-2\lambda\eta_{11}\nabla_{\overline{z}}\nabla_{z} +\lambda\left(-\frac{i\eta_{11}}{2\tau_2}F_{12} +\frac{1}{2}\eta_{ij}g_{ij}\right).
\label{eq: antihol V}
\end{align}

We remark that the first term in Eq.~\eqref{eq: hol V} is minimized for holomorphic sections of the complex line bundle $L\to \BZ^2$ whose fiber at $\bf{k}$ is the image of $P(\bf{k})$, i.e., Bloch wave functions $\ket{u_{\bf{k}}}$ living in the Chern band  [$P(\textbf{k})\ket{u_{\bf{k}}}=\ket{u_{\bf{k}}}$] and satisfying
\begin{align*}
\nabla_{\overline{z}}\ket{u_{\bf{k}}}=0.
\end{align*}
The vector space of (global) solutions of the above equation is denoted $H^0(\BZ^2,L)$. If the Chern number $\mathcal{C}$ of $L$ is positive then the above equation has $\mathcal{C}$ linearly independent solutions by the Riemann-Roch theorem. If the Chern number of $L$ is negative, then, the above equation has no (global) solutions and we have to turn to Eq.~\eqref{eq: antihol V}, in which the first term is minimized by antiholomorphic sections of $L$, i.e., those sections satisfying
\begin{align*}
\nabla_{z}\ket{u_{\bf{k}}}=0,
\end{align*}
which will have $|\mathcal{C}|$ linearly independent solutions as it corresponds to Bloch wave functions on the Chern band, i.e., sections of $L$, which satisfy the constraint that they are holomorphic with respect to the opposite complex structure on the Brillouin zone.

The second term in Eqs.~\eqref{eq: hol V} and~\eqref{eq: antihol V} vanishes  when
\begin{align*}
\mp iF_{12}=\eta^{11}\tau_2\eta_{ij}g_{ij}=\sqrt{\det(\eta^{-1})}\eta_{ij}g_{ij}=\eta_{ij}g_{ij},
\end{align*}
where the last equality follows from unimodularity of $\eta$. This identity holds when the triple of structures $(g,-iF/2=\omega, \pm j_{\tau})$ satisfies $\omega(\cdot,\pm j_{\tau}\cdot)=g$, so that
\begin{align*}
g_{ij}(\bf{k})=f(\bf{k})\eta^{ij},
\end{align*}
for some non-negative smooth function $f(\bf{k})$, and
\begin{align*}
\eta_{ij}g_{ij}=2f(\bf{k}),
\end{align*}
and also
\begin{align*}
\mp\frac{iF_{12}}{2}=\sqrt{\det (g)}= f(\bf{k}).
\end{align*}
We note that, for any Chern insulator, an inequality $\sqrt{\det (g)} \ge |F_{12}/2|$ holds~\cite{roy:14}, which follows from the Cauchy-Schwarz inequality; the saturation of the inequality is related to the Bloch states being holomorphic or anti-holomorphic as described in Refs.~\cite{oza:mer:21:phys:published,mer:oza:21:math:published}. 
We thus see that vanishing of the expectation value of $\overline{V}(\bf{k})$ is equivalent to the condition that we can choose local Bloch states that are holomorphic or anti-holomorphic functions (depending on whether the sign of the first Chern number is negative or positive, respectively).

The guiding-center orbitals of the Chern band are then defined to be a basis of the Hilbert space of Bloch wave functions on the Chern band, i.e., sections of $L$ [a Hilbert subspace of the total single-particle Hilbert space of the system defined by those Bloch states $\ket{u_{\bf{k}}}$ which satisfy $P(\bf{k})\ket{u_{\bf{k}}}=\ket{u_{\bf{k}}}$ for all $\bf{k}\in\BZ^2$] composed of eigenvectors of the confining potential $\overline{V}(\bf{k})$, with $\eta$ variationally chosen to optimize the dispersive second term of Eqs.~\eqref{eq: hol V} and~\eqref{eq: antihol V} (see Refs.~\cite{claa:lee:tho:qi:dev:15,lee:cla:tho:17}). Observe that if the band is K\"{a}hler with respect to some flat K\"{a}hler structure with complex structure $j_{\tau}$ ($-j_{\tau}$), then the second term in Eq.~\eqref{eq: hol V} (Eq.~\eqref{eq: hol V}) is naturally minimized and the conformal factor $f(\textbf{k})$ appearing above is simply a constant.

The challenge of obtaining a microscopic description of fractional quantum Hall states associated with fractionally filled Chern bands reduces, according to~\cite{claa:lee:tho:qi:dev:15}, to determining the deformed guiding-center orbitals, above defined as eigenstates of $\overline{V}(\textbf{k})$, upon placing a fractional quantum Hall liquid on the lattice. Given an appropriate $\eta$, any fractional Chern insulator can in principle be captured by Laughlin-like many-body trial ground states, constructed from the single-body eigenstates of Eq.~\eqref{eq: projected confining potential}.

The discussion above refers to the choice of ``guiding-center'' bases in the conventional fractional quantum Hall effect and in fractional Chern insulators. In the latter case, to ensure the stability with respect to interactions one further needs that the Berry curvature is uniform~\cite{roy:14,jac:mol:roy:15} to ensure that the resulting projected density algebra is isomorphic to the $W_{\infty}$-algebra found in the ordinary FQHE. For $\tau=i$ it is not just isomorphic but exactly the same algebra.

The construction in the present manuscript, see Sec.~\ref{sec: The Bergman kernel prescription}, provides asymptotically flat K\"{a}hler bands and, hence, it provides a way to obtain optimal Chern bands for hosting fractional Chern insulators, for arbitrary anisotropy as described by the modular parameter $\tau\in\mathcal{H}$. The fact that our formalism allows for arbitrary anisotropy $\tau$, allows to account for scenarios in which, due to symmetry reasons, it will be more favourable to form a fractional Hall fluid on the lattice if the geometry of the Chern band is anisotropic. Intuitively, this should be the case for lattices that are not conformal to the square lattice.

\section{General expressions for (quasi-)K\"ahler bands from holomorphic maps from $\BZ^2$ to $\mathbb{C}P^{n-1}$}
\label{sec: Quasi-Kahler two-band models and holomorphic maps from BZ2 to CP1}
We now explain how K\"ahler and quasi-K\"ahler bands can be generally expressed as holomorphic maps from $\BZ^2$ to $\mathbb{C}P^{n-1}$, using meromorphic functions on complex tori. We first consider the case where $n = 2$, which are two-band models and thus one can only construct quasi-K\"ahler bands. We then extend the argument to $n \ge 2$ and present a general expression for a map $\BZ^2 \to \mathbb{C}P^{n-1}$, which any K\"ahler and quasi-K\"ahler band should obey.

For two-band models a quasi-K\"{a}hler band is determined by a holomorphic map $P:\BZ^2\to \mathbb{C}P^1$, or equivalently, because we can identify $\mathbb{C}P ^1\cong \mathbb{C}\cup\{\infty\}$ (Riemann sphere), by a meromorphic function over a torus equipped with the structure of a complex manifold. Holomorphic maps, unlike smooth maps, are very ``rigid'', something that is ultimately related to the fact that holomorphic functions are Taylor series expandable everywhere. In the following, for illustration purposes, we describe meromorphic functions over a complex torus and we will see how they are parametrized by their zeros and poles.
For simplicity, we equip the Brillouin zone $\BZ^2$ with the $2\pi\left(\mathbb{Z}+i\mathbb{Z}\right)-$periodic complex coordinate $z=k_1+ik_2$, making it a complex torus.
We note that generalization to a general case of $z = k_1 + \tau k_2$ with $\tau \in \mathcal{H}$ is straightforward, as we will see below.
Observe that we are using coordinates $(k_1,k_2)$ such that the action of the reciprocal lattice by translations is given by shifts by integer multiples of $2\pi$ on each coordinate. Let
\begin{align*}
\theta(z,\tau=i):=\theta(z)=\sum_{n\in\mathbb{Z}}e^{-\pi n^2 +2\pi i nz},
\end{align*}
be the associated \emph{theta function}.

Observe that
\begin{align*}
\theta(z+m)=\theta(z), \text{ for } m\in\mathbb{Z},
\end{align*}
and
\begin{align*}
\theta(z+mi)&=\sum_{n\in\mathbb{Z}}e^{-\pi n^2 +2\pi in z -2\pi mn}\\
&=\sum_{n\in\mathbb{Z}}e^{-\pi (n+m)^2 +2\pi i(n+m) z +\pi m^2-2\pi i mz}\\
&=e^{\pi m^2 -2\pi imz}\theta(z), \text{ for all } m\in\mathbb{Z}.
\end{align*}

The $\theta$ function has a unique simple zero at $1/2+i/2 \mod \mathbb{Z}+i\mathbb{Z}$~\cite{mir:95}. 
We define the translated theta functions by
\begin{align*}
\theta^{(x)}(z)=\theta(z-(1/2+i/2) -x),\ x\in\mathbb{C},
\end{align*}
which has zeros at $x+\mathbb{Z}+i\mathbb{Z}$.
A holomorphic map $R:\BZ^2\to \mathbb{C}P^1$, i.e., a meromorphic function on $\BZ^2$, can be constructed by taking products and ratios of the translated theta functions
\begin{align}
\label{eq: R(z)}
    R(z)=\frac{\prod_{i}\theta^{(x_i)}\left(\frac{z}{2\pi}\right)}{\prod_{j}\theta^{(y_j)}\left(\frac{z}{2\pi}\right)}
\end{align}
for a set of complex numbers $x_1, x_2, \cdots, x_N$ and $y_1, y_2, \cdots, y_N$ satisfying the condition
\begin{align}
    \sum_{i=1}^{N}x_i-\sum_{j=1}^{N}y_j\in\mathbb{Z}. \label{eq:condition_meromorphic}
\end{align}
One can explicitly confirm that $R(z+2\pi i) = R(z)$ holds if and only if the condition Eq.~(\ref{eq:condition_meromorphic}) is satisfied. Note that in Eq.~\eqref{eq: R(z)} we performed a rescaling on the $z$ variable to ensure the correct periodicity properties under translations by reciprocal lattice vectors.

Actually, every meromorphic function on the complex torus $\mathbb{C}/\mathbb{Z}^2$ is, \emph{up to a multiplicative constant}, of this form (see, for instance, Proposition~2.7 and Lemma~3.14 of~\cite{mir:95}).

For a given holomorphic map $R:\BZ^2\to \mathbb{C}P^1$, we can construct a two-band momentum-space Hamiltonian $H(\mathbf{k}) = - \varepsilon n(\mathbf{k}) \cdot \sigma$, where $\varepsilon$ is a positive constant, $\sigma = (\sigma_1, \sigma_2, \sigma_3)$ are the Pauli matrices and $n$ is a map $n:\BZ^2\to S^2\subset\mathbb{R}^3$ with
\begin{align*}
n(z)=\frac{1}{1+|R(z)|^2}\left(2\mbox{Re} \left(R(z)\right),2\mbox{Im} \left(R(z)\right), 1-|R(z)|^2\right).
\end{align*}
The eigenvalues of the Hamiltonian are $\pm \varepsilon$, and thus flat over the Brillouin zone, and the eigenvectors are, up to normalization, $ (-\overline{R(z)},1)$ and $(1, R(z))$. Therefore, the lower band of this Hamiltonian is described by the map
$f:\BZ^2\to \mathbb{C}P^1$ given by
\begin{align*}
f:\BZ^2\to \mathbb{C}P^1;\ z\mapsto f(z)=\left[\prod_{j}\theta^{(y_j)}\!\left(\frac{z}{2\pi}\right)\!:\!\prod_{i}\theta^{(x_i)}\!\left(\frac{z}{2\pi}\right)\!\right],
\end{align*}
in homogeneous coordinates.
The lowest band of this Hamiltonian is, by construction, quasi-K\"ahler with respect to the flat complex structure determined by $\tau = i$. We note that, since this is a two-band Hamiltonian, there must be points in the Brillouin zone where $\det (g) = 0$ due to Theorem~3 of Ref.~\cite{mer:oza:21:math:published}. That means that the Berry curvature will be always non-negative. The reason is that, given that $R:\BZ^2\to \mathbb{C}P^1$ is holomorphic, it will necessarily be orientation preserving at every point. This follows from the fact that, locally, $R$ is described by a holomorphic map $z\mapsto w(z)$, where $z=k_1+i k_2$ is the holomorphic coordinate on the Brillouin zone and $w$ is a holomorphic coordinate on the sphere. The orientations on both manifolds are locally determined by the $2-$forms $dk_1\wedge dk_2=\frac{i}{2}dz\wedge d\bar{z}$ and $\frac{i}{2}dw\wedge d\bar{w}$, respectively. The pullback of $\frac{i}{2}dw\wedge d\bar{w}$ then reads $\frac{i}{2}|\frac{\partial w}{\partial z}|^2 dz\wedge d\bar{z}$. Since $|\frac{\partial w}{\partial z}|^2\geq 0$, it follows that the map will be orientation preserving on each point where the derivative does not vanish, i.e., away from the points where the map is not an immersion or, equivalently, $\det(g)=0$.

We can generalize the above construction of a holomorphic map to the case of $\mathbb{C}P^{n-1}$ by:
\begin{widetext}
\begin{align*}
f:\BZ^2\to \mathbb{C}P^{n-1};\ z\!\mapsto \left[c_1\prod_{i_1}\theta^{(x^1_{i_1})}\left(\frac{z}{2\pi}\right):c_2\prod_{i_2}\theta^{(x^2_{i_2})}\left(\frac{z}{2\pi}\right)\!:\!\cdots \!:\! c_n\prod_{i_n}\theta^{(x^n_{i_n})}\left(\frac{z}{2\pi}\right)\right],
\end{align*}
\end{widetext}
where $c_1,\dots,c_n\in\mathbb{C}$, with at least one of the $c_i$'s different from zero,  $1\leq i_j\leq N$, for all $j=1,\dots,n$, and
\begin{align}
\sum_{j=1}^{N}x^{i}_{j}-x^{k}_j\in\mathbb{Z}, \text{ for all } i<k.
\label{eq: condition on zeros}
\end{align}
To see that this is completely general, note that to have a general well-defined holomorphic map $f:\BZ^2\to \mathbb{C}P^{n-1}; z\mapsto [f_1(z):\dots:f_{n}(z)]$ we need a collection of functions $f_i: \mathbb{C}\to\mathbb{C}$ depending holomorphically in $z$ and which satisfy
\begin{align}
f_i(z+G)=e_{G}(z)f_i(z), \text{ for all } i\in \{1,\dots,n\},
\end{align}
where $G=G_1+iG_2$ represents a reciprocal lattice vector and $e_{G}(z)$ is a holomorphic multiplier which is the same for all functions. This is the case because the projective space does not care about the overall scale of the vector. The equation above tells us that the $f_i(z)$'s define $n$ sections of a holomorphic line bundle over the complex torus. Note, however that we can write 
\begin{align}
[f_1(z):\dots:f_{n}(z)] &=\left[1:\frac{f_2(z)}{f_1(z)}:\dots:\frac{f_n(z)}{f_1(z)}\right] \nonumber \\
&=[1:R_1(z):\dots:R_{n}(z)],
\end{align}
where the $R_i(z)$'s are now meromorphic functions over the complex torus. Since any meromorphic function over the complex torus can be written, up to a multiplicative constant, as a ratio of theta functions we understand that the prescription is general.

Thus, any K\"ahler and quasi-K\"ahler band, having a flat complex structure defined by $\tau=i$, can be written in this form, and can be expressed by appropriately choosing the values of $c_1, \cdots, c_n$ and $x_j^i$ to satisfy Eq.~(\ref{eq: condition on zeros}). The discussion above is easily generalized to a complex torus $\mathbb{C}/\left(\mathbb{Z}\oplus \tau\mathbb{Z}\right)$, for some $\tau\in\mathcal{H}$, by replacing $\theta(z)=\theta(z,\tau=i)$ with
\begin{align}
\theta(z,\tau)=\sum_{n\in\mathbb{Z}}e^{i\pi \tau n^2 +2\pi i nz},
\end{align}
where now $z=k_1+\tau k_2$. In fact, since any torus with the structure of a complex manifold, i.e., a Riemann surface of genus $1$, is biholomorphic to a complex torus $\mathbb{C}/\left(\mathbb{Z}\oplus \tau\mathbb{Z}\right)$, for some $\tau\in\mathcal{H}$, see Proposition~5.2 of~\cite{mir:95}, it follows that any K\"{a}hler and quasi-K\"{a}hler band, after composing the map with a suitable map giving the biholomorphism, can be described in this way \footnote{To be precise, it means that if we equip the Brillouin zone with the structure of a complex manifold, denoted $(\BZ^2,j)$, with $j$ being the complex structure, there exists a biholomorphism (i.e., an isomorphism of complex manifolds) $\varphi:(\BZ^2,j)\to \mathbb{C}/\left(\mathbb{Z}\oplus \tau\mathbb{Z}\right)$ such that
$
d\varphi\circ j=j_{\tau}\circ d\varphi,    
$
with $j_{\tau}$ the uniform flat complex structure associated with $\tau\in\mathcal{H}$. Note that, since $\varphi$ is in particular a diffeomorphism, we also have
$
j\circ d\varphi^{-1}=d\varphi^{-1}\circ j_{\tau} .
$
This means that if $P:(\BZ^2,j)\to \mathbb{C}P^{n-1}$ is a holomorphic map, meaning
$
dP\circ j=J_{FS}\circ dP,
$
where $J_{FS}$ is the Fubini-Study complex structure, then, $\tilde{P}=P\circ \varphi^{-1}: \mathbb{C}/\left(\mathbb{Z}\oplus \tau\mathbb{Z}\right)\to \mathbb{C}P^{n-1}$ is holomorphic with respect to the flat complex structure $j_{\tau}$ defined by $\tau\in\mathcal{H}$, i.e.,
\begin{align*}
d\tilde{P}\circ j_{\tau}&=dP\circ d\varphi^{-1}\circ j_{\tau} = dP\circ j\circ d\varphi^{-1}\\
&=J_{FS}\circ dP\circ d\varphi^{-1} =J_{FS}\circ d\tilde{P},
\end{align*}
where we used the chain rule for differentials $d\tilde{P}=d(P\circ \varphi^{-1})=dP\circ d\varphi^{-1}$. Now the projector $\tilde{P}$ will have a description in terms of meromorphic functions on a complex torus parametrized by a complex coordinate $z=k_1+\tau k_2$ as described in the main text.
}.

Because of the number of free parameters at hand, tuning these parameters to obtain the desired (quasi-)K\"ahler band is generally a nontrivial task. As discussed in Sec.~\ref{sec: Background on fractional Chern insulators and band geometry}, in the context of the fractional Chern insulators, one is interested in finding bands with the quantum metric and the Berry curvature which are as uniform as possible over the Brillouin zone. One approach taken by Lee \emph{et al.}~\cite{lee:cla:tho:17} was to set an ansatz wave function in terms of the Weierstrass zeta functions, instead of the theta functions, and to flatten the geometrical properties by tuning a collection of variational parameters. In the next section, we present an alternative approach, for which one needs to determine only an auxiliary Hermitian holomorphic line bundle, for a fixed value of the modular parameter $\tau$, and total number of bands. After the explicit construction of the holomorphic line bundle for a given $\tau$, which is made possible using the theory of theta functions, the only parameter in our approach becomes the number of total bands. Despite the simplicity of our approach, we show that, in an appropriate asymptotic limit of large total number of bands, our method gives rise to the desired flat quantum metric and Berry curvature, satisfying $|F_{12}|/2=\sqrt{\det(g)}$. Our approach is inspired by the one by Kovrizhin \emph{et al.}~\cite{kov:dou:moe:13}, in which skyrmion lattices with desired properties were constructed using theta functions.

\section{The Bergman kernel prescription}
\label{sec: The Bergman kernel prescription}
We want to build a holomorphic map $f:\BZ^2\to \mathbb{C}P^{n-1}$, for some $n>1$, which induces a flat geometrical structure over the Brillouin zone $\BZ^2$. We first show that K\"ahler bands which are geometrically flat are solely determined by the modular parameter $\tau$ and the first Chern number $-\mathcal{C}$, with $\mathcal{C}>0$ (the minus sign coming from the relation between our convention of the Berry curvature and the pullback of the Fubini-Study form, as we will see more explicitly below). Flat K\"ahler bands are characterized by a compatible triple, $(\omega, j_{\tau}, g)$, where each of the structures is uniform over the Brillouin zone. Let us denote the modular parameter of the system by $\tau$, which means that the complex variable $z = k_1 + \tau k_2$ gives a complex coordinate on the Brillouin zone satisfying $d s^2 = \sum_{i,j}g_{ij}dk_i dk_j = g_{11} |dz|^2$. From this relation, we can deduce $\tau = (g_{12} + i\sqrt{\det g})/g_{11}$. The almost complex structure $j_{\tau}$ satisfies $j_{\tau}(\partial_z) = i\partial_z$ and $j_{\tau}(\partial_{\bar{z}}) = -i\partial_{\bar{z}}$, from which we can deduce
\begin{align}
    j_{\tau} = 
    \frac{1}{\mathrm{Im}(\tau)}
    \begin{pmatrix}
    -\mathrm{Re}(\tau) & -|\tau|^2 \\ 1 & \mathrm{Re}(\tau)
    \end{pmatrix}
    =
    \frac{1}{\sqrt{\det (g)}}\begin{pmatrix} -g_{12} & -g_{22} \\ g_{11} & g_{12} \end{pmatrix}
\end{align}
in the basis of $\partial/\partial k_1$ and $\partial/\partial k_2$.
Since $g_{ij} = \sum_{k}\omega_{ik}(j_{\tau})_{j}^{k}$ from the compatibility of the K\"ahler structure, we can then show $\omega = \sqrt{\det (g)}dk_1 \wedge dk_2$. On the other hand, the negative of the Chern number of the K\"ahler band is $\mathcal{C} = \int_{\BZ^2}\omega/\pi =4\pi \sqrt{\det (g)}$.
Therefore, for a given value of the modular parameter $\tau$ and of the Chern number $-\mathcal{C}$, the flat K\"ahler structure we want to obtain is
\begin{align}
 \omega&=\sqrt{\det (g)}dk_1\wedge dk_2=\frac{\mathcal{C}}{4\pi}dk_1\wedge dk_2=\frac{\mathcal{C}}{4\pi}\frac{1}{\bar{\tau}-\tau}dz\wedge d\overline{z}, \nonumber\\
 j_{\tau} &= 
    \frac{1}{\mathrm{Im}(\tau)}
    \begin{pmatrix}
    -\mathrm{Re}(\tau) & -|\tau|^2 \\ 1 & \mathrm{Re}(\tau)
    \end{pmatrix},\
    g =\frac{\mathcal{C}}{4\pi \mathrm{Im}(\tau)} |dz|^2,
\label{eq: flat Kahler triple}
\end{align}
which is the K\"{a}hler structure that we want the map $f$ to induce on $\BZ^2$ (up to a constant scaling on $\omega$ and $g$). Hence, the triple $(\omega, j_{\tau},g)$ associated with a K\"{a}hler band is completely determined by $\tau$ and $-\mathcal{C}$, as claimed.
We note the symplectic form $\omega$ can be written in terms of a positive function $h_*$ as $\omega = -(i/2)\partial \bar{\partial} \log h_*$, where
\begin{align}
    h_*=\exp\left(\frac{\mathcal{C}}{4\pi}\frac{i}{\tau-\overline{\tau}}(z-\overline{z})^2\right)=\exp\left(-\frac{\mathcal{C}}{2\pi}\mathrm{Im}(\tau) k_2^2\right), \label{eq:hstar}
\end{align}
and $\partial=dz\frac{\partial}{\partial z}\wedge\cdot$, $\overline{\partial}=d\overline{z}\frac{\partial}{\partial \overline{z}}\wedge\cdot$ are the Dolbeault operators. With this function $h_*$, the K\"ahler potential $K$, satisfying $\omega = i\partial\bar{\partial}K$, can be written as
\begin{align}
    K = -\frac{1}{2}\log h_* = \frac{\mathcal{C}}{4\pi}\mathrm{Im}(\tau) k_2^2 \text{ or } h_*=e^{-2K}.
\end{align}

In order to build the map $f$, we are inspired by the idea of the \emph{K\"{a}hler quantization}. The theory of geometric quantization~\cite{woo:97,hal:03} gives us a prescription to quantize a K\"{a}hler manifold $(\BZ^2,\omega,J,g)$ provided the symplectic form $\omega$ satisfies the quantization condition, namely that $\omega/2\pi$ represents an integer cohomology class -- $[\omega/2\pi]\in H^2(\BZ^2;\mathbb{Z})$ or, equivalently, that $\int_{\BZ^2} \omega$ is $2\pi$ times an integer. If this is the case, then $-i\omega$ represents the 1st Chern class of a Hermitian holomorphic line bundle $L\to \BZ^2$ equipped with the Chern connection [i.e., the unique connection whose $(0,1)$ part coincides with the Cauchy-Riemann operator of $L$]. Once we are given the data $L\to \BZ^2$, known as the \emph{pre-quantum line bundle}, then, the K\"{a}hler quantization of $\BZ^2$ is defined as the space $H^0(\BZ^2,L)\subset \Gamma(\BZ^2,L)$ (equipped with the $L^2$-norm and completed with respect to it) consisting of (square-integrable) global holomorphic sections of $L$. From the physics point of view, this prescription appears naturally in the physics of fermions, in two spatial dimensions, in the presence of an external uniform magnetic field, i.e., in the quantum Hall effect, where the lowest Landau level is precisely described by $H^0(M,L)$ where $M$ is the surface corresponding to the physical sample~\cite{klev:16}. For example, if $M=\mathbb{R}^2$, and $L=\mathbb{R}^2\times\mathbb{C}$ with magnetic field represented by the Faraday $2-$form $F=-idx\wedge dy$ we recover the space of (square-integrable) holomorphic functions on the plane with measure $e^{-|z|^2/4}d^2x$. The difference here, is that $M$ is not the real space physical sample, but rather the quasi-momentum space. This geometric quantization procedure appears also naturally in the discussion of Sec.~\ref{sec: Background on fractional Chern insulators and band geometry}, where the confining potential $\overline{V}(\bf{k})$ isolates $H^0(\BZ^2,L)$, where $L\to \BZ^2$ is the line bundle associated to the Chern band. There, the complex structure over $\BZ^2$ was singled out by the variational flat metric $\eta$, much like here we will fix it \emph{a priori} below. Finally, for the construction below, it will be convenient and, in fact, crucial to take not $H^0(\BZ^2,L)$ but rather $H^0(\BZ^2,L^{\otimes p})$, for some integer $p>0$, which, as we will soon see, determines the total number of bands in the system.

We are now in condition to build the map $f:\BZ^2\to\mathbb{C}P^{n-1}$. For that, let us consider an auxiliary Hermitian holomorphic line bundle $L\to \BZ^2$, which, at this point, is arbitrary besides having nontrivial Chern number $\mathcal{C}>0$, which is also known as the degree of $L$ (note that this auxiliary holomorphic line bundle $L$ will have positive Chern number $\mathcal{C}>0$, unlike the resulting K\"{a}hler band which will have negative Chern number $-p\mathcal{C}$). Note that the fact that $L$ is a holomorphic line bundle means that its transition functions are holomorphic with respect to the complex structure specified by $\tau$ in the Brillouin zone. More specifically, they will depend holomorphically on the complex variable $z=k_1+\tau k_2$. The fact that $L$ is Hermitian means that it comes equipped with a Hermitian metric on the fibers, which on a given (holomorphic) gauge is represented by positive function $h>0$. The Chern connection on $L$ has the property that, in the local holomorphic gauges, the local gauge field and the associated curvature are given by, respectively,
\begin{align}
A=\partial \log h  \text{ and } F=-\partial \overline{\partial}\log h,
\end{align}
where $h$ is the representative of the Hermitian metric in this gauge. Observe how the $(0,1)$ part of $A$ vanishes, and hence a local holomorphic gauge is holomorphic with respect to the connection $\nabla$, i.e., it is annhihilated by the covariant derivative $\nabla_{\frac{\partial}{\partial \bar{z}}}$. This is the local form of the defining property of the Chern connection, namely, that it is the unique connection whose $(0,1)$ part coincides with the Cauchy-Riemann operator of $L$ (see Sec.~6 of Ref.~\cite{che:67} and Sec.~4 of Ref.~\cite{kob:14}, in particular, Proposition~(4.9)).

Borrowing the ideas from the K\"ahler quantization, we choose the line bundle $L$ so that the Hermitian metric $h$ coincides with the positive function $h_*$ defined in Eq.(\ref{eq:hstar}), which is related to the K\"ahler potential of the flat K\"ahler band we want to achieve. Taking $h = h_*$, the curvature of the line bundle is
\begin{align}
    F &= -\partial \bar{\partial} \log h = -2i\omega = -i\frac{\mathcal{C}}{2\pi}dk_1 \wedge dk_2.
    \label{eq:curvature}
\end{align}
A line bundle $L$ with such curvature $F$ will have first Chern number
\begin{align}
\int_{\BZ^2}\frac{iF}{2\pi}=\mathcal{C}. 
\end{align}

Now suppose for a moment that such Hermitian holomorphic $L$ exists and we have built it. An explicit construction, where, without loss of generality, $\mathcal{C}=1$, will be given below in Sec.~\ref{sec: Explicit construction}. We now take $p$ copies of the line bundle $L$, where the number $n$ in $\mathbb{C}P^{n-1}$ is related to the Chern number through $n = p\mathcal{C}$. The fact that $\dim H^0(\BZ^2,L^{\otimes p})=\deg (L^{\otimes p})=p\mathcal{C}=n$ follows from the celebrated Riemann-Roch theorem~\cite{mir:95,huy:05}.
We collect a basis of $H^0(\BZ^2,L^{\otimes p})$, and call them $\{s_j\}_{j=1}^{p\mathcal{C}}$.
We take the map $f:\BZ^2\to\mathbb{C}P^{n-1}$ given by
\begin{align*}
f:\BZ^2\to\mathbb{C}^{n-1}; \; z\mapsto [a_1(z):\dots:a_{p\mathcal{C}}(z)],
\end{align*}
with $a_{j}(z)$ being the holomorphic components of $s_j$,  $j=1,\dots,p\mathcal{C}$, in a holomorphic gauge $s$ [in which gauge the $(0,1)$ part of the holomorphic gauge field representing the Chern connection, which is compatible with the Hermitian metric defined by the $p-$th power of $h$, $h^p$, vanishes].
Note that the $a_{j}$'s will satisfy appropriate boundary conditions so as to define sections of $L^{\otimes p}$, more precisely, the section $s_j$ as determined by $s_j=s a_j$ will be periodic, but $s$ and $a_j$ will satisfy
\begin{widetext}
\begin{align}
\label{eq: transformation rules}
s(\textbf{k}+\textbf{G})=s(\textbf{k})\left(e_{\textbf{G}}(z)\right)^{-1} \text{ and } a_{j}(z+G_1+\tau G_2)=e_{\textbf{G}}(z)a_j(z), \text{ for all } j=1,\dots,p\mathcal{C}, 
\end{align}
\end{widetext}
for a given \emph{system of holomorphic multipliers} $\{e_{\textbf{G}}(z)\}$, where $\textbf{G}=(G_1,G_2)$ is an arbitrary element of the reciprocal lattice; i.e., a collection of functions $\{e_{\textbf{G}}(z)\}$, labeled by the reciprocal lattice, depending holomorphically in $z$ and that satisfy
\begin{align}
e_{\textbf{G}+\textbf{G}'} (z)=e_{\textbf{G}'}(z+G_1+\tau G_2)e_{\textbf{G}}(z),
\end{align}
where $\textbf{G}=(G_1,G_2)$ and $\textbf{G}'=(G'_1,G'_2)$ are arbitrary reciprocal lattice vectors written in the $(k_1,k_2)-$coordinates. The holomorphic multipliers are enough to reconstruct $L^{\otimes p}$ completely~\cite{mum:74}. The transformation rules of Eq.~\eqref{eq: transformation rules} ensure that the map to the projective space is well-defined (provided $s_1,\dots,s_{p\mathcal{C}}$ do not vanish simultaneously) because, for all $z$,
\begin{widetext}
\begin{align}
[a_1(z+G_1+\tau G_2):\dots:a_{p\mathcal{C}}(z+G_1+\tau G_2)]=[e_{\textbf{G}}(z)a_1(z):\dots:e_{\textbf{G}}(z)a_{p\mathcal{C}}(z)]=[a_1(z):\dots:a_{p\mathcal{C}}(z)],
\end{align}
\end{widetext}
since the equivalence class does not care about an overall scale. Furthermore the map $f$ is holomorphic in $z$, by construction. The pullback of the Fubini-Study symplectic form under $f$ is
\begin{align*} 
f^*\omega_{FS}=\frac{i}{2}\partial\overline{\partial}\log\left(\sum_{j}|a_j(z)|^2\right).
\end{align*}
Furthermore
\begin{align*}
p\omega -f^*\omega_{FS}&=-\frac{i}{2}\partial\overline{\partial} \log h^p -\frac{i}{2}\partial\overline{\partial}\log\left(\sum_j|a_j|^2\right)\\
&=-\frac{i}{2}\partial \overline{\partial} \log B,
\end{align*}
with $B$ being the (diagonal) of the so-called \emph{Bergman kernel}~\cite{zel:98}:
\begin{align}
\label{eq: Bergman kernel}
B=\sum_{j} h^p|a_j|^2=\sum_j h^p(s_j,s_j),
\end{align}
provided $\{s_j\}_{j=1}^{p\mathcal{C}}$ form an orthonormal basis with respect to the $L^2$-inner product induced by $h$ and $\omega$ on $\Gamma(\BZ^2,L^{\otimes p})$, i.e.,
\begin{align}
\langle s_i, s_j\rangle_{L^2} &=\int_{\BZ^2} h^p(s_i,s_j)\; \omega \nonumber\\
&=\frac{\mathcal{C}}{4\pi}\int d^2k\; e^{-\frac{p\mathcal{C}\tau_2}{2\pi}k_2^2} \overline{ a_i(z)}a_j(z) =\delta_{ij}.
\label{eq: L^2 inner product}
\end{align}
In Eqs.~\eqref{eq: Bergman kernel} and~\eqref{eq: L^2 inner product}, $h^{p}(s_i,s_j)=h^{p}\overline{a}_i a_j$ denotes the evaluation of the Hermitian metric $h^{p}$ on the sections $s_i$ and $s_j$ determined by the functions $a_i$ and $a_j$ -- it is, therefore, a periodic function. For large $p$, it is known that there is an asymptotic expansion assuming the form~\cite{zel:98}
\begin{align*}
B=p +A_1p^0 +A_{2}p^{-1}+\dots+ A_{k}p^{1-k}+\dots,
\end{align*}
where the $A_i$'s are smooth functions. In particular, as $p$ becomes large, we see that $f^*\omega_{FS}$ goes to $p\omega$, meaning that the symplectic structure, and consequently, by compatibility, the K\"{a}hler structure, becomes flat.

From a general consideration, one can show that it is not possible to construct a K\"ahler band which is geometrically completely flat, that is, $f^* \omega_{FS}$ is a constant over the entire Brillouin zone, with a model with a finite number of bands. We prove this no-go theorem in Appendix~\ref{sec: no-go theorem}. Allowing an infinite number of bands, it is possible to construct bands which are geometrically completely flat; Landau levels are such examples.

We note that the diagonal of the Bergman kernel also appears, in real space rather than in momentum space, in the context of the lowest Landau level over a Riemann surface as the particle density for the associated many-particle integer Hall effect state~\cite{klev:16} and it has been previously approached through path integrals in Ref.~\cite{dou:klev:10}.
\section{Explicit construction}
\label{sec: Explicit construction}
We now provide an explicit construction of geometrically nearly flat K\"ahler bands along the prescription described in the previous section. For this purpose, we first need to specify the auxiliary Hermitian holomorphic line bundle $L \to \mathrm{BZ}^2$, whose curvature satisfies Eq.~(\ref{eq:curvature}). A line bundle over $\BZ^2$ is uniquely characterized by specifying a system of holomorphic multipliers, Eq.~(\ref{eq: transformation rules}). We define the line bundle $L$ to have the holomorphic multipliers, denoted $\{e^{L}_{\bf{G}}(z)\}$, by
\begin{align*}
e^{L}_{\bf{G}=(2\pi m,2\pi n)}(z)=e^{-i\pi\tau n^2 -2\pi i n\left(\frac{z}{2\pi}\right)},
\end{align*}
where $\bf{G}$ is in the reciprocal lattice, so that $m,n\in\mathbb{Z}$. Observe that for a gauge field $A$ to be consistent with the above system of holomorphic multipliers and to define a connection on $L$, it must satisfy
\begin{align}
A(\bf{k}+\bf{G}) -A(\bf{k})=-\left(e^{L}_{\bf{G}}\right)^{-1}de^{L}_{\bf{G}}=-d\log e^{L}_{\bf{G}},
\end{align}
for all $\bf{G}$ in the reciprocal lattice, because the holomorphic multipliers define the holomorphic transition functions of the line bundle $L$. We note that the $(1,0)$ form
\begin{align}
A=2\pi i k_2 dz=\partial \log h=\partial \log\exp\left(\frac{1}{4\pi}\frac{i}{\tau-\overline{\tau}}(z-\overline{z})^2\right)
\label{eq: gauge field for L}
\end{align}
does satisfy this property and is, indeed, a sensible gauge field for the bundle $L$. It is also clear that the prescribed gauges are holomorphic since $A$ is a $(1,0)-$form. Furthermore, from Eq.~(\ref{eq: gauge field for L}) , we can read off a Hermitian metric [for which $A$ is representing the Chern connection] as described by the positive function $h=e^{\frac{1}{4\pi}\frac{i}{\tau-\overline{\tau}}(z-\overline{z})^2}=e^{-\frac{\mathrm{Im}(\tau) k_2^2}{2\pi}}$. The associated curvature is
\begin{align}
F=-\partial\overline{\partial}\log h=2\pi i dk_2\wedge dz=-2\pi i dk_1\wedge dk_2,    
\end{align}
thus, Eq.~\eqref{eq:curvature}, with $\deg (L)=\mathcal{C}=\int_{\BZ^2}idA/2\pi=1$, is satisfied by $L$ equipped with this connection. 

The holomorphic multipliers for $L^{\otimes p}$ are simply the $p-$th power of those of $L$, namely,
\begin{align}
&e^{L^{\otimes p}}_{2\pi(m,n)}(z) \nonumber \\
&=\left(e^{L}_{2\pi(m,n)}(z)\right)^p= e^{-\pi i\tau pn^2 - in pz}, \text{ for } m,n\in\mathbb{Z}.
\end{align}
The holomorphic sections of such line bundle $L^{\otimes p}$ can be explicitly given by theta functions as we proceed to describe below.

For $\tau\in\mathcal{H}$, we define the theta function with characteristics $a$ and $b$ by
\begin{align*}
\vartheta \left[\begin{array}{c}
a\\
b
\end{array}
\right](z,\tau )=\sum_{n\in \mathbb{Z}}e^{ i\tau\pi(n+a)^2 +2\pi i (n+a)(z+b)}, \text{ for } a,b\in\mathbb{R}.
\end{align*}
This function satisfies
\begin{align*}
\vartheta \left[\begin{array}{c}
a\\
b
\end{array}
\right](z\!+\! m\!+\!ni,\tau)\! = \!e^{-i\pi\tau n^2 \!-\! 2\pi i n z \!+\!2\pi i(am \!-\! b n)}\vartheta \left[\begin{array}{c}
a\\
b
\end{array}
\right](z,\tau),
\end{align*}
for $m,n\in\mathbb{Z}$. The relation to the standard theta function
\begin{align*}
\theta(z,\tau)=\sum_{n\in\mathbb{Z}}e^{\pi in^2\tau +2\pi inz},
\end{align*}
is
\begin{align*}
\vartheta \left[\begin{array}{c}
a\\
b
\end{array}
\right](z,\tau )=e^{\pi i a^2\tau +2\pi i a(z+b)}\theta(z+a\tau +b,\tau).
\end{align*}

The functions
\begin{align}
a_{j}(z) &=\vartheta\left[\begin{array}{c}
\frac{j}{p}\\
0
\end{array}
\right]\left(p\frac{z}{2\pi},p \tau \right)\\
&= e^{\pi i\tau \frac{j^2}{p} + i jz} \theta\left(p\frac{z}{2\pi} +j\tau, p\tau\right) ,\ j=0,\dots,p-1,
\label{eq: aj's}
\end{align}
define a basis of holomorphic sections $\{s_{j}\}_{j=0}^{p-1}$ of a line bundle $L^{\otimes p}\to\BZ^2$. The proof that these functions do satisfy the periodicity required by the holomorphic multipliers is given in Appendix~\ref{sec: additional properties of theta functions}. We note that, in real space, the same theta functions with characteristics arise when looking at the lowest Landau level over the torus, see Refs.~\cite{ono:01,klev:16}.

A natural gauge field consistent with this system of multipliers of $L^{\otimes p}$  is simply $p A$, with $A$ as in Eq.~\eqref{eq: gauge field for L}.  It immediately follows that the associated curvature is $pF=-ip(dk_1\wedge dk_2)/(2\pi)$ and that the first Chern number of $L^{\otimes p}$ is $p$.

Furthermore we can show that (see Ref.~\cite{klev:16} for an analogous formula)
\begin{align*}
\langle s_i, s_j\rangle_{L^2} &=\frac{1}{(2\pi)^2}\int_{\BZ^2} e^{-p \frac{ k_2^2}{2\pi} } \overline{a_i}a_j dk_1\wedge dk_2 \\
&=\sqrt{\frac{i}{p(\tau-\overline{\tau})}} \delta_{ij},
\end{align*}
hence, the map we want is induced by the vector
\begin{align*}
z\mapsto (a_0(z),\dots,a_{p-1}(z)),
\end{align*}
for large $p$, namely,
\begin{align*}
f_p:\BZ^2\to\mathbb{C}P^{p-1}; \; z\mapsto [a_0(z):\dots:a_{p-1}(z)],
\end{align*}
or, in terms of orthogonal rank $1$ projectors, a smooth map $z\mapsto P_p(z)$ with
\begin{align*}
\bra{i}P_p(z)\ket{j}=\frac{a_i(z)\overline{a_j(z)}}{\sum_{k}|a_k(z)|^2},
\end{align*}
with $\ket{i}$, $i=0,\dots,p-1$, the canonical basis of $\mathbb{C}^p$. Observe that, since $\theta(z,\tau)$ has zeros in $\frac{1}{2}(1+\tau) +\mathbb{Z}+\tau\mathbb{Z}$, it follows that $a_j$, or equivalently, $s_j$ has zeros at positions 
\begin{widetext}
\begin{align}
\frac{2\pi}{p}\left(\frac{1}{2}+r\right) +2\pi\left(\frac{1}{2}-\frac{j}{p}\right)\tau +2\pi\mathbb{Z}+2\pi\tau\mathbb{Z},\ r=0,\dots,p-1,
\end{align}
\end{widetext}
for $j=0,\dots,p-1$. Thus, the sections do not vanish simultaneously and the map $f_p$ is well-defined. For large enough $p$ (actually for $p>2$), the band is K\"{a}hler. In the large $p$ limit the band is K\"{a}hler and actually the K\"{a}hler structure is asymptotically flat $(f_{p}^*\omega_{FS}, j_{\tau},f_{p}^*g_{FS})\sim (p\, \omega,j_{\tau},p\, g)$. Having built an approximately flat K\"{a}hler band (for finite $p$) one can always build a tight-binding model with a flat dispersion by declaring the momentum-space Hamiltonian to be $H(\textbf{k})=I_p-2P_p(\textbf{k})$, $\bf{k}\in\BZ^2$ and then inverse Fourier transforming it to obtain the model in the lattice, namely, $H(\bf{r}_i,\bf{r}_j)=\int_{\BZ^2}\frac{d^2k}{(2\pi)^2}e^{i\bf{k}\cdot\left(\bf{r}_i-\bf{r}_j\right)} H(\bf{k})$ with $p$ orbitals/internal degrees of freedom per site. We can always build a tight-binding model with a flat dispersion with this prescription, but it will always have long range hoppings provided its Chern number is nontrivial~\cite{chen:maza:sei:tan:14}. We can make the tight-binding model strictly local by truncating the hoppings. This truncation procedure will, however, always violate (even if weakly) the flat structures. In particular, we will see that the resulting Chern bands will not, in general, be K\"{a}hler bands.

\subsection{Numerical results}
\label{subsec: Results}
\begin{figure}[h!]
\begin{center}
\subfigure[$\tau = i$]{
\includegraphics[width= 0.48 \textwidth]{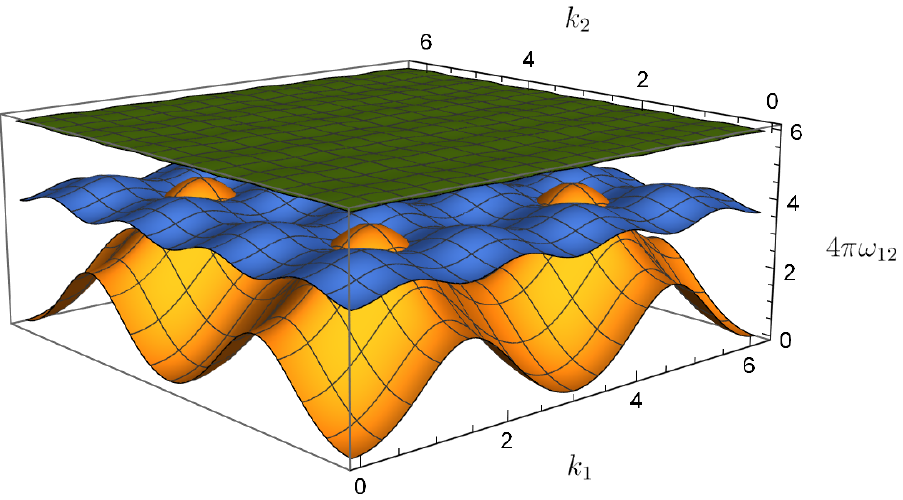}}
\subfigure[$\tau=e^{i\pi/3}$]{
\includegraphics[width= 0.48 \textwidth]{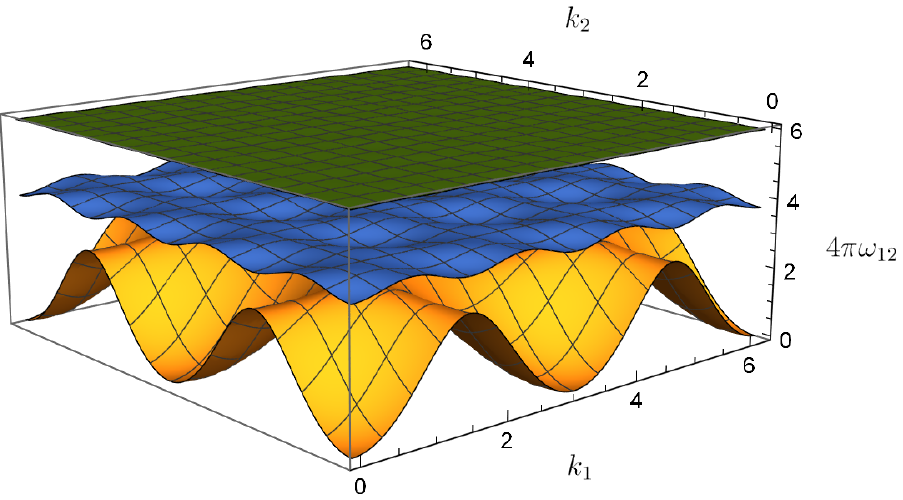}}
\subfigure[$\tau=2e^{i\pi/7}$]{
\includegraphics[width= 0.48 \textwidth]{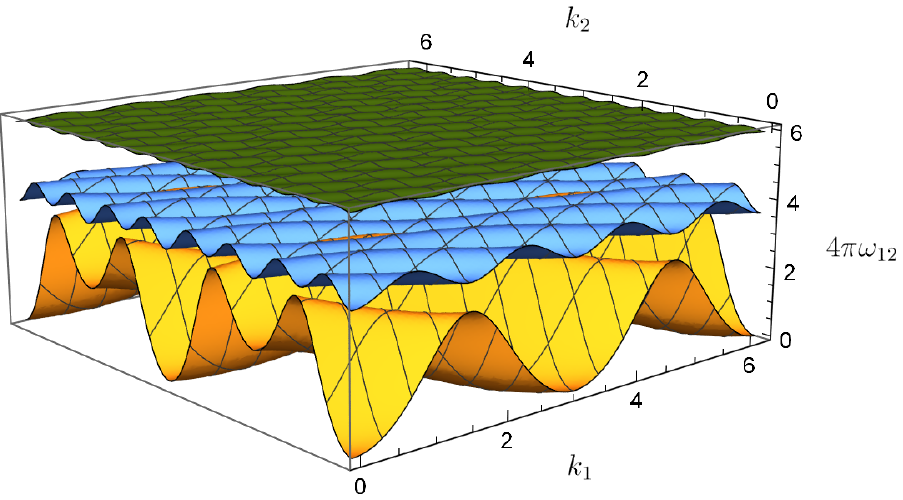}}
\caption{Plot of $4\pi \omega_{12}$, where $\omega_{12}$ the only independent component of the pullback $f_p^*\omega_{FS}$, as a function of the quasimomentum $\bf{k}\in \BZ^2$, for $p=2$ (orange), $p=4$ (blue), $p=6$ (green), for various anisotropies $\tau$. We have numerically confirmed that $4\pi \sqrt{\det(g)}$, representing the quantum volume form, takes the same value as $4\pi\omega_{12}$ for all the cases shown here.}
\label{fig: tau} 
\end{center}
\end{figure}

We now perform numerical simulation of the constructed model. We consider the isotropic case $\tau=i$, and also anisotropic cases $\tau=e^{i\pi/3}$ and $\tau=2e^{i\pi/7}$, and show how the Berry curvature and the quantum volume form flatten for large $p$. 
By construction, our model is K\"ahler and thus $ \sqrt{\det(g (\bf{k}))} = |F_{12}(\bf{k})|/2 = \omega_{12}(\bf{k}) $ always holds, where $\omega_{12}dk_1\wedge dk_2 =(-i/2)F_{12}dk_1\wedge dk_2$ is the pullback of the Fubini-Study symplectic form, and we have also numerically confirmed this equality.
In Figs.~\ref{fig: tau}(a),~\ref{fig: tau}(b), and~\ref{fig: tau}(c), we plot $4\pi \omega_{xy}(\bf{k})$, for $p=2,4,6$, and for $\tau=i$, $e^{i\pi/3}$, and $2e^{i\pi/7}$, respectively.
We see that already for $p=6$, $4\pi \omega_{xy}(\bf{k})$ reaches the flat value of $4\pi \omega= 4 \pi \sqrt{\det(g)}dk_1\wedge dk_2=p\,dk_1\wedge dk_2$ very well.

Our model of the K\"ahler band is constructed in momentum space. When interpreted as a real-space lattice model, it contains long-range hoppings at any length.
We now consider the effect of truncating the hopping to obtain short-range models. We consider how our momentum space model can be translated into a square lattice model in real space, and analyze the consequence of truncation of hopping on the geometrical structure in momentum space.

The real-space tight-binding model associated to $H(\bf{k})=I_{p}-2P_p(\bf{k})$ is
\begin{align}
H(\bf{r}_i,\bf{r}_j)=\int_{\BZ^2}\frac{d^2k}{(2\pi)^2} H(\bf{k})e^{i\bf{k}\cdot\left(\bf{r}_i-\bf{r}_j\right)},
\end{align}
where $\bf{r}_i,\bf{r}_j\in\mathbb{Z}^2$ label the positions on the lattice. The truncation of the hoppings can be done by taking the function $f_R(\bf{r})=f_R(x_1,x_2)$ defined by
\begin{align}
&f_R(x_1,x_2)=\begin{cases}
1,  \text{ if } -\lfloor R\rfloor \leq x_i \leq \lfloor R\rfloor,\ i=1,2,\\
0, \text{ otherwise.}
\end{cases},
\end{align}
for some $R$ and considering the truncated real space Hamiltonian $H_{R}(\bf{r}_i,\bf{r}_j)=H(\bf{r}_i,\bf{r}_j)f(\bf{r}_i-\bf{r}_j)$. It is convenient, to simplify notation, to write $B_{R}$ for the set of translations of the lattice $\bf{r}$ with $f_{R}(\bf{r})=1$. Note that $f_{R}(\bf{r})$ is just the indicator function of the set $B_{R}$. Since under Fourier transformation the product becomes a convolution and the Fourier coefficients of $f_{R}(\bf{r})$ are, evidently,
\begin{align}
\widetilde{f}_{R}(\bf{k})=\sum_{\bf{r}\in B_{R}}e^{-i\bf{k}\cdot \bf{r}},
\end{align}
we finally arrive at the truncated Hamiltonian
\begin{align}
H_{R}(\bf{k})&=\int_{\BZ^2} \frac{d k'}{(2\pi)^2} H(\bf{k}-\bf{k}')\widetilde{f}_{R}(\bf{k}')\nonumber\\
&=\int_{\BZ^2} \frac{d k'}{(2\pi)^2}\sum_{\bf{r}\in B_{R}}H(\bf{k}-\bf{k}')e^{-i\bf{k}'\cdot \bf{r}}.
\label{eq: truncated tight binding model}
\end{align}
Note that, as $R\to\infty$, we approach the original Hamiltonian we started with, because $\widetilde{f}_{R}(\bf{k})$ approaches the Dirac delta distribution on the Brillouin zone $(2\pi)^2\delta^2(\bf{k})$. Now observe that a Riemann sum for the above expression is provided by
\begin{align}
\frac{1}{N_1N_2}\sum_{\bf{n}\in \mathbb{Z}_{N_1}\times \mathbb{Z}_{N_2}}\sum_{\bf{r}\in B_R}H(\bf{k} -\bf{k}'_{\bf{n}})e^{-i\bf{k}'_{\bf{n}}\cdot \bf{r}},
\label{eq: truncated tight binding model finite}
\end{align} 
with $\bf{k}'_{\bf{n}}=(2\pi n_1/N_1,2\pi n_2/N_2)$ with $\bf{n}=(n_1,n_2)\in \mathbb{Z}_{N_1}\times \mathbb{Z}_{N_2}$, where $\mathbb{Z}_{N}$ denote the integers modulo $N$. Provided $R<\min(N_1,N_2)$, the above expression corresponds to the tight-binding model one would get by truncating the one in the finite system with $N_1$ and $N_2$ sites in the $x_1$ and $x_2$ directions, respectively, with periodic boundary conditions. In the limit when $N_1$, $N_2$ are large, the above expression is a good approximation for the Bloch Hamiltonian obtained by truncating the tight-binding model in $\mathbb{Z}^2$.

\begin{widetext}

\begin{figure}[h!]
\centering
\subfigure[$R=\infty$ vs. $R=1$]{
\includegraphics[width= 0.48 \textwidth]{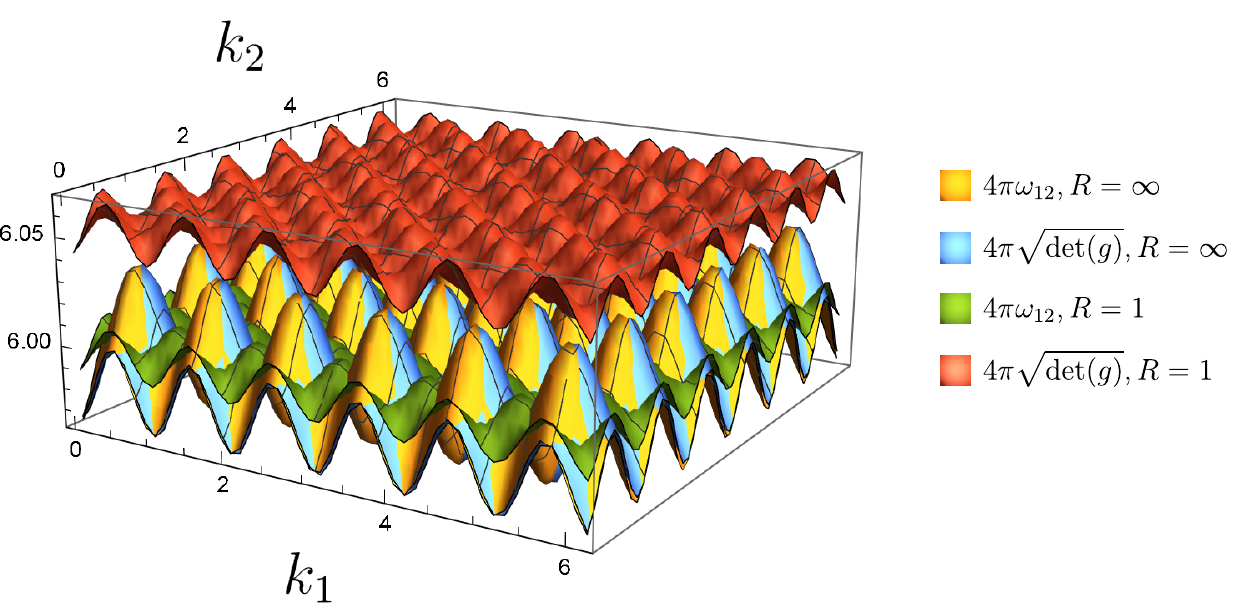}
\includegraphics[width= 0.48 \textwidth]{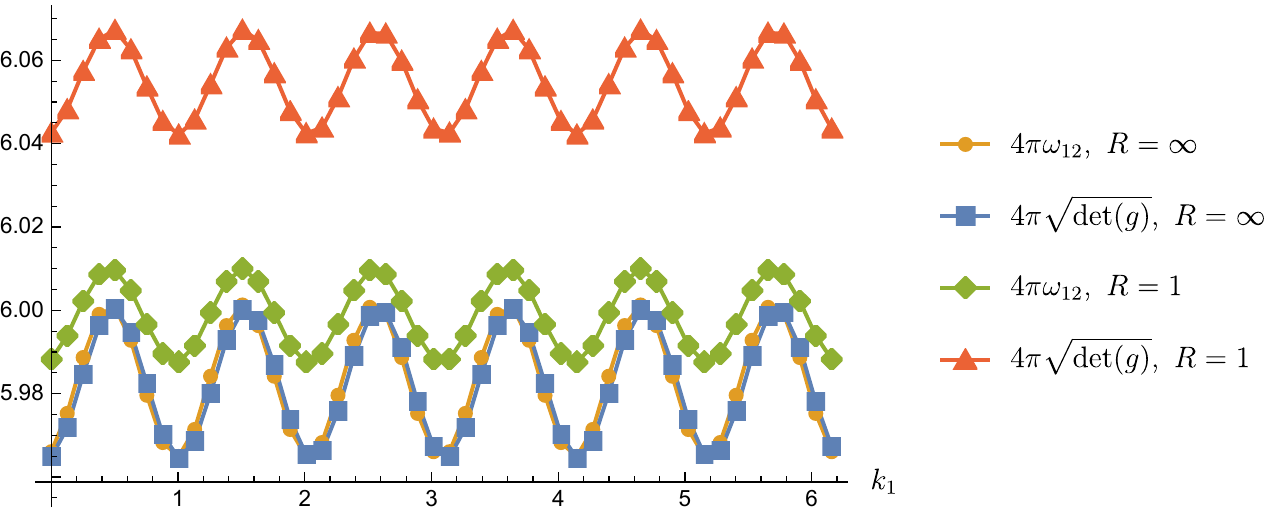}
}
\subfigure[$R=\infty$ vs. $R=2$]{
\includegraphics[width= 0.48 \textwidth]{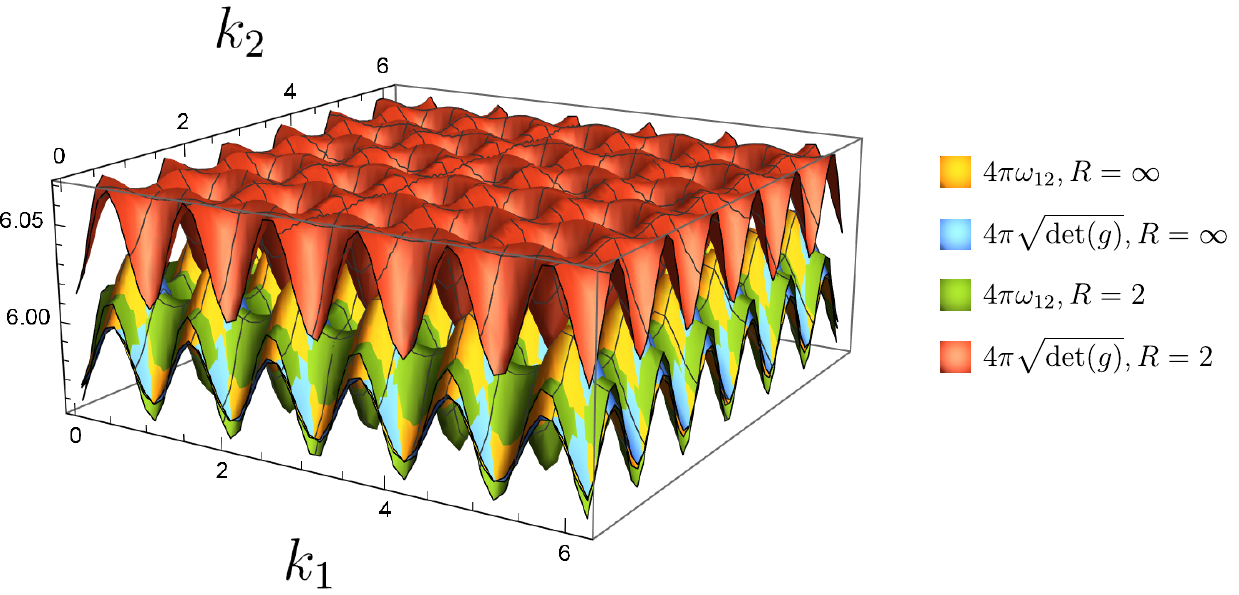}
\includegraphics[width= 0.48 \textwidth]{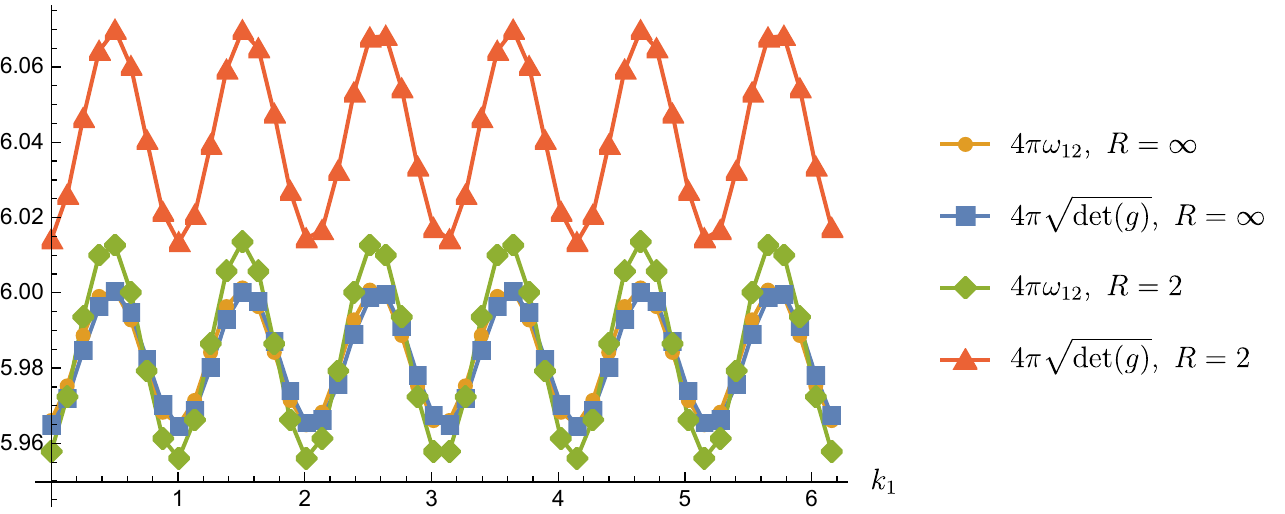}
}
\subfigure[$R=\infty$ vs. $R=3$]{
\includegraphics[width= 0.48 \textwidth]{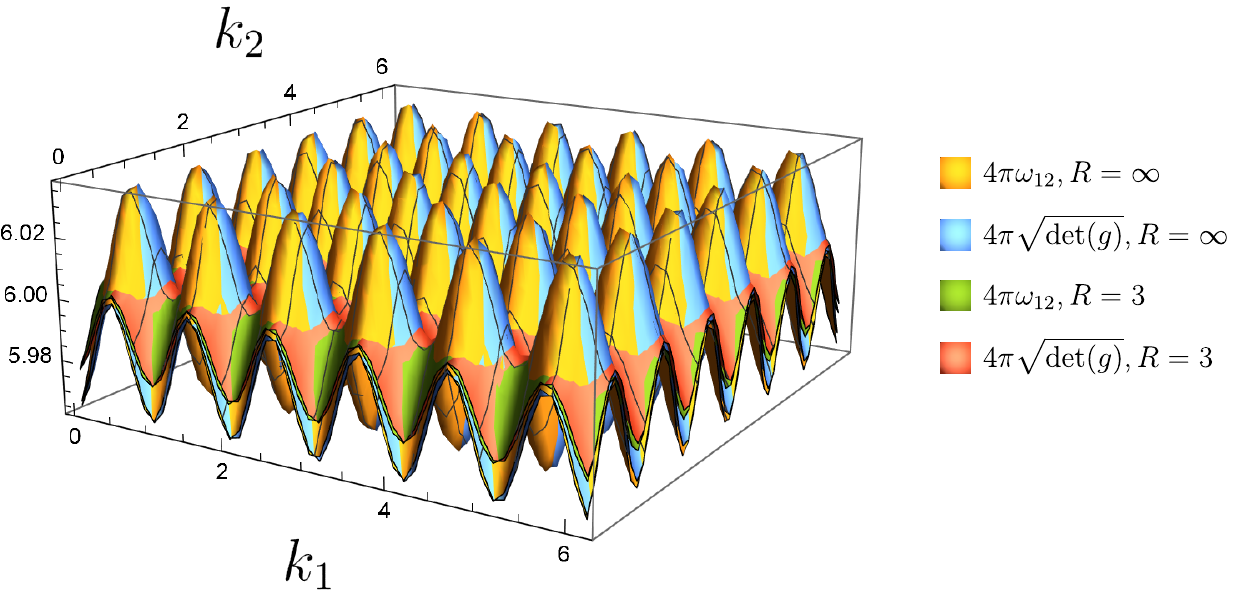}
\includegraphics[width= 0.48 \textwidth]{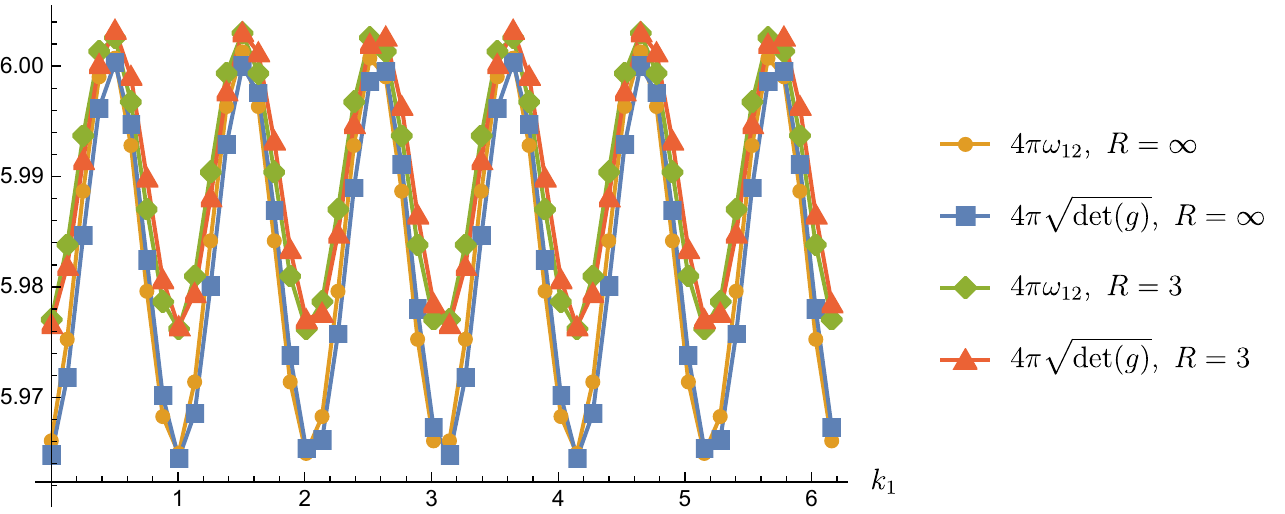}
}
\caption{The geometrical structure, $4\pi\sqrt{\det(g)}$ and $4\pi\omega_{12}$, of truncated models as a function of $\bf{k}\in \BZ^2$, for $p=6$. In the top panel, $R=1$ is compared to the long range case $R=\infty$; in the middle panel $R=2$ is compared to the $R=\infty$ case, and in the bottom panel $R = 3$ is compared to $R = \infty$. On the left column we present the variation of the geometric quantities along the whole of $\BZ^2$ and on the right column, for clarity, we present the variation along the cut at $k_2=0$. On the right column, the markers represent the data points and the lines are there for a clear visualization of the data.}
\label{fig: truncatedtb}
\end{figure} 
\end{widetext}

\begin{figure}
    \centering
    \subfigure[Relative fluctuations of $\omega_{12}$]{
\includegraphics[width= 0.48 \textwidth]{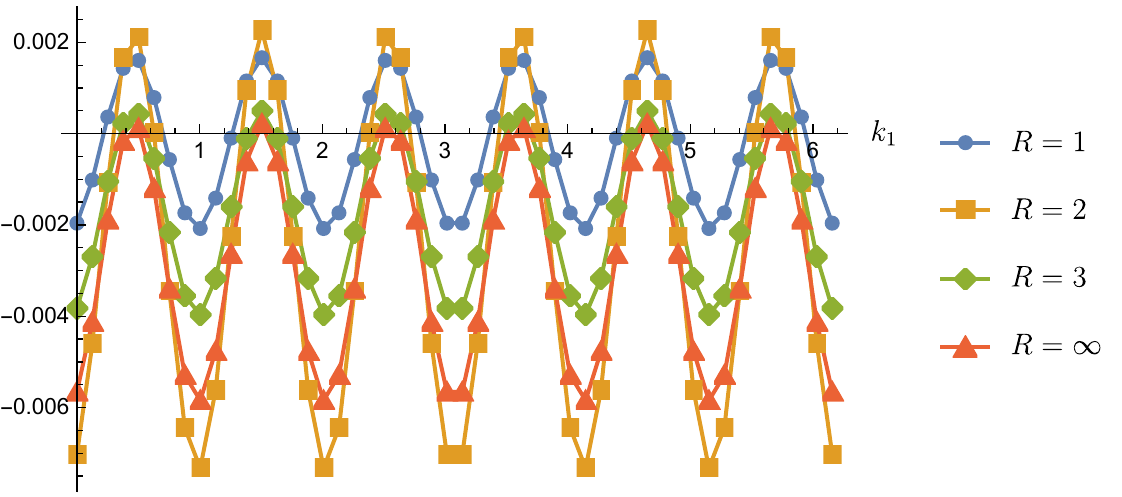}
}
\subfigure[Relative fluctuations of $\sqrt{\det(g)}$]{
\includegraphics[width= 0.48 \textwidth]{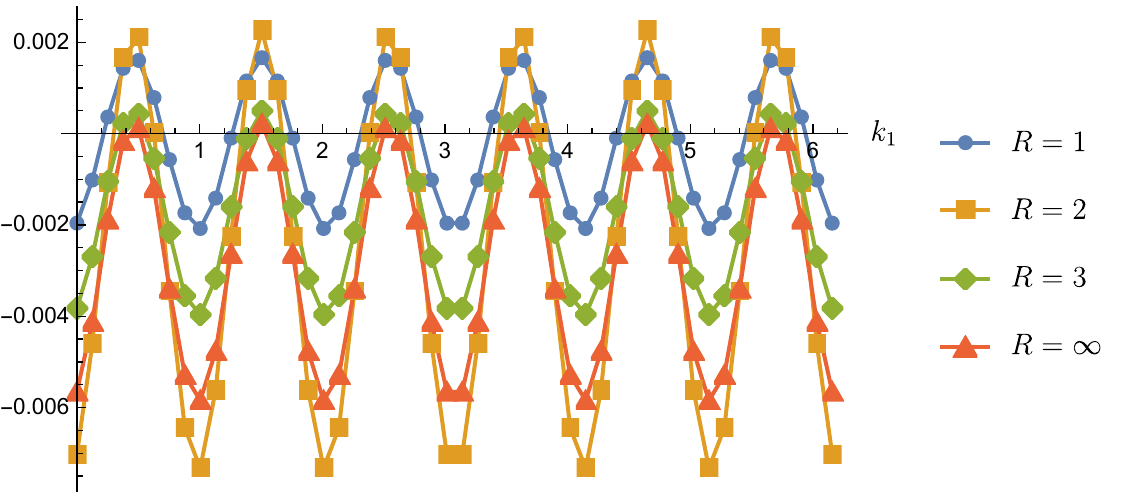}
}
    \caption{The relative fluctuations, for $p=6$, of $\omega_{12}$ and $\sqrt{\det(g)}$ as measured from the flat value $p/4\pi$, as a function of $k_1$ along the cut $k_2=0$. The markers represent the data points and the lines are there for a clear visualization of the data.}
    \label{fig: relative fluctuations}
\end{figure}

In our simulation, we take a square lattice with $N_1=N_2=50$ sites and periodic boundary conditions on the $x_1$ and $x_2$ directions, we fix $\tau=i$ and we take $p=6$, and consider how the quantum geometry of the lowest energy band of the Hamiltonian is affected as we truncate the allowed hoppings by shrinking $R$. In Fig.~\ref{fig: truncatedtb}, we present the numerically calculated $\sqrt{\det(g)}$ and $\omega_{12}$ for $R=1$, $R=2$, and $R=3$, corresponding to allowing up to first, second, and third nearest neighbor hoppings, respectively, and compared them to the $R=\infty$ long-range case, where there is no truncation. Additionally, in  Fig.~\ref{fig: relative fluctuations}, we present the relative fluctuations of $\sqrt{\det(g)}$ and $\omega_{12}$ with respect to the flat value $p/4\pi$ (with $p=6$), for $R=1$, $R=2$, and $R=3$, and $R=\infty$, along the submanifold defined by $k_2=0$.
We can see that when truncating the hopping at $R = 3$, namely, including up to third-nearest-neighbor hoppings, the geometrical structure is already almost identical to the long-range case of $R = \infty$.
When $R = 1$ and $R=2$, we can see the effect of truncation more clearly.
The effect of truncation is two-fold. The first effect is that $\sqrt{\det(g)}$ and $\omega_{12}$ are no longer equal, implying the breaking of holomorphicity of the Chern bands. However, the difference between $\sqrt{\det(g)}$ and $\omega_{12}$ is not so large even for the case of nearest-neighbor model of $R=1$; in our simulation the difference is around one percent as one can read off from Fig.~\ref{fig: truncatedtb}. The second effect is that the flatness of the geometrical quantities, $\sqrt{\det(g)}$ and $\omega_{12}$, will change. What we have numerically observed is that the geometrical quantities will not become more dispersive, and sometimes they can become even {\it flatter} when truncating the hopping. We note that, even when $R = 1$ and $R=2$, the Chern number, $-\int_{\BZ^2}\omega/\pi =-6$, is the same as $R = \infty$, and thus the bands are adiabatically connected to the ideal K\"ahler band. We have also numerically checked the cases with smaller $p$, such as $p = 3$ or $p=4$, and the overall behavior remains the same. We note that $p=2$ is special; the band cannot be K\"ahler because of the constraints that $\det (g(\mathbf{k})) = 0$ should hold somewhere in the Brillouin zone, and thus bands cannot be made geometrically flat.

\section{Conclusions}
\label{sec: Conclusions}

In this manuscript, we described a systematic method to obtain K\"ahler bands, whose metric and curvature satisfies the equality $\sqrt{\det (g(\mathbf{k}))} = |F_{12}(\mathbf{k})|/2$, become flat in momentum space as the total number of bands is increased.
We also showed that it is not possible to obtain a perfectly flat K\"ahler band with a finite number of total bands.
We provided an explicit construction of our method using theta functions, and we have numerically observed that we can achieve fairly flat K\"ahler bands with a reasonably small number of total bands. We have also shown that truncating the lattice model to include only local hoppings will not affect the flatness of the bands, but does make the band deviate from the perfect K\"ahler bands, leading to $\sqrt{\det (g(\mathbf{k}))} \ge |F_{12}(\mathbf{k})|/2$. 

The Bergman kernel prescription can be applied to the more general case in which one replaces the flat K\"ahler structure $(\omega, g, j_{\tau})$ in Eq.~\eqref{eq: flat Kahler triple} by some arbitrary K\"ahler structure $(\omega',g',j')$. In that case, the resulting map $f_p$, built from an orthonormal basis of holomorphic sections of the $p$-th tensor power of an auxiliary Hermitian holomorphic (with respect to the new complex structure $j'$) line bundle $L'$ satisfying the quantization condition $-\partial\overline{\partial}\log h'=-2i\omega'$, will also yield, asymptotically in $p$, the compatible triple $(f_p^*\omega_{FS},j',f_p^*g_{FS})\sim (p\omega',j',pg')$. This prescription would then allow us to engineer K\"ahler bands with a prescribed profile of Berry curvature and quantum metric which may be potentially interesting for engineering of Chern bands with desired properties.
Such generalizations of our method will be discussed in more detail in future works.

Upon submission of this work, we became aware of a related no-go theorem which was very recently presented in Ref.~\cite{varjas:abouelkomsan:yang:bergholtz:21}. They showed that it is impossible to have a flat K\"ahler band coming from a lattice system with a finite number of lattice sites per unit cell (hence with a finite total number of bands), which agrees with our no-go theorem.

\begin{acknowledgments}
B.M. acknowledges very stimulating discussions with J. P. Nunes and J. M. Mour\~{a}o. T.O. acknowledges support from JSPS KAKENHI Grant Number JP20H01845, JST PRESTO Grant Number JPMJPR19L2, JST CREST Grant Number JPMJCR19T1, and RIKEN iTHEMS. B.M. acknowledges the support from SQIG -- Security and Quantum Information Group, the Instituto de Telecomunica\c{c}\~oes (IT) Research Unit, Ref. UIDB/50008/2020, funded by Funda\c{c}\~ao para a Ci\^{e}ncia e a Tecnologia (FCT), European funds, namely, H2020 project SPARTA, as well as  projects QuantMining POCI-01-0145-FEDER-031826 and PREDICT PTDC/CCI-CIF/29877/2017.
\end{acknowledgments}

\appendix
\section{Additional properties of theta functions}
\label{sec: additional properties of theta functions}
We had
\begin{align*}
a_{j}(z)&=\vartheta\left[\begin{array}{c}
\frac{j}{p}\\
0
\end{array}
\right]\left(p\frac{z}{2\pi},p \tau \right)\\
&= e^{\pi i\tau \frac{j^2}{p} + i jz} \theta(p\frac{z}{2\pi} +j\tau, p\tau) ,\ j=0,\dots,p-1,
\end{align*}
We can then write
\begin{align*}
a_{j}(z)=e^{\pi i\tau \frac{j^2}{p} + i jz} \sum_{n\in\mathbb{Z}}e^{\pi i\tau pn^2 +2\pi i n\left(\frac{p z}{2\pi}+j\tau\right)}
\end{align*}
We have, for $m,n\in\mathbb{Z}$,
\begin{align*}
&a_{j}(z+2\pi m +2\pi n\tau )\\
&=e^{\pi i\tau\frac{j^2}{p} +ij(z+2\pi m +2\pi n\tau)}\theta(p\frac{z}{2\pi} + pm + \tau pn +j\tau, p\tau)\\
&=e^{\pi i\tau\frac{j^2}{p} \!+\! ij(z+2\pi m +2\pi n\tau )\!-\!\pi i\tau p n^2 \!-\! 2\pi i n\left(p\frac{z}{2\pi}+j\tau\right)}\theta(p\frac{z}{2\pi}\!+\!j\tau, pi)\\
&=e^{-\pi i\tau pn^2 - in pz -2\pi i nj\tau +2\pi ij n\tau +2\pi ij m }a_{j}(z)\\
&=e^{-\pi i\tau pn^2 - in pz}a_{j}(z),
\end{align*}
which agrees with the system of holomorphic multipliers, $\{e_{\textbf{G}}(z)\}$, for the holomorphic line bundle $L^{\otimes p}$,
\begin{align}
e_{2\pi(m,n)}(z)= e^{-\pi i\tau pn^2 - in pz}, \text{ for } m,n\in\mathbb{Z}.
\end{align}

The associated gauge field is given by
\begin{align*}
A_p &=pA=ip k_2 \frac{dz}{2\pi}=\partial \log \exp\left(- p\frac{\tau_2 k_2^2}{2\pi}\right)\\
&=\frac{-i}{2}\frac{\partial}{\partial k_2}\left[\log \exp\left(- p\frac{\tau_2 k_2^2}{2\pi}\right)\right] dz,
\end{align*}
where $p=\deg (L^{\otimes p})$ is the first Chern number of $L^{\otimes p}$.
\section{Formulae for performing Numerics}
Let $\ket{\Psi(\bf{k})}$ be the unit vector in $\mathbb{C}^{p}$ defined by
\begin{align}
\ket{\Psi(\bf{k})}=\frac{(a_0(z),\dots,a_{p-1}(z))}{\sqrt{\sum_{j}|a_j(z)|^2}},
\end{align}
with the $a_j$'s defined by Eq.~\eqref{eq: aj's}, $z=k_1+\tau k_2$, and $\bf{k}\in \BZ^2$. The discussion below allows us to numerically evaluate the quantum metric and the Berry curvature in terms of $\ket{\Psi(\bf{k})}$. Additionally, we will also use the formulas below to compute the quantum metric and the Berry curvature for the case when $\ket{\Psi(\bf{k})}$ is not defined by the above formula but rather defined as the lowest energy eigenstate of the short-range tight-binding model specified by the truncation of the Bloch Hamiltonian $H(\bf{k})=I_p-2P_p(\bf{k})$, denoted  $H_{R}(\bf{k})$ and defined in Eq.~\eqref{eq: truncated tight binding model}.

To compute the \emph{quantum metric}, we used the fidelity between neighboring states
\begin{align*}
\left|\langle \Psi(\bf{k})| \Psi(\bf{k}+ \varepsilon v)\rangle\right| = 1-\frac{1}{2} g(v,v)\varepsilon^2 +\cdots,
\end{align*}
where $v=(v^1,v^2)$ defines a tangent vector $\sum_i v_i\frac{\partial}{\partial k_i}\in T_{\bf{k}}\BZ^2$ and $\varepsilon$ is a small number. 

To compute the \emph{Berry curvature}, we used the fact that for a given path $\bf{k}(t)$, $(0\leq t\leq 1)$, we have, for large $N$,
\begin{align*}
\prod_{i=0}^{N-1}\langle \Psi(\bf{k}(\frac{i+1}{N}))|\Psi(\bf{k}(\frac{i}{N}))\rangle \approx \exp\left(-\int_{0}^{1} A\left(\frac{d \bf{k}}{dt}\right)dt\right),
\end{align*}
In particular, if we take a loop that is a boundary of a surface $\Sigma\subset \BZ^2$, we have
\begin{align*}
\prod_{i=0}^{N-1}\langle \Psi(\bf{k}(\frac{i+1}{N}))|\Psi(\bf{k}(\frac{i}{N}))\rangle\approx \exp(-\int_{\Sigma}F).
\end{align*}
We can take $\Sigma$ to be an infinitesimal rectangle with vertices $\bf{k}_1\equiv \bf{k}_5=\bf{k}$, $\bf{k}_2=\bf{k}+\varepsilon_1 \hat{\bf{x}}_1$, $\bf{k}_3=\bf{k}+\varepsilon_1\hat{\bf{x}}_1+\varepsilon_2\hat{\bf{x}}_2$ and $\bf{k}_4=\bf{k}+\varepsilon_2\hat{\bf{x}}_2$ in an orientation consistent with that of the standard one in $\BZ^2$, for small $\varepsilon_1$ and $\varepsilon_2$. Here $\hat{\bf{x}}_1=(1,0)$ and $\hat{\bf{x}}_2=(0,1)$. This then gives
\begin{align*}
-i\text{Im}\log\left[\prod_{i=1}^{4}\langle \Psi(\bf{k}_{i+1})|\Psi(\bf{k}_i)\rangle\right]\approx F_{12}(\bf{k})\varepsilon_1\varepsilon_2.
\end{align*}
Using this notation we also see that
\begin{align*}
-\log\left[|\langle \Psi(\bf{k}_{2})|\Psi(\bf{k}_1)\rangle|^2\right] \approx & g_{11}(\bf{k})\varepsilon_1^2,\\
 -\log\left[|\langle \Psi(\bf{k}_{4})|\Psi(\bf{k}_1)\rangle|^2\right]\approx & g_{22}(\bf{k})\varepsilon_2^2,\\
-\log\left[|\langle \Psi(\bf{k}_{3})|\Psi(\bf{k}_1)\rangle|^2\right]\approx & g_{11}(\bf{k})\varepsilon_1^2 +2g_{12}(\bf{k})\varepsilon_1 \varepsilon_2 \\
&+g_{22}(\bf{k})\varepsilon_2^2.
\end{align*}
This justifies why we have to take the imaginary part in the previous equation:
\begin{align*}
&-\text{Re}\log\left[\prod_{i=1}^{4}\langle \Psi(\bf{k}_{i+1})|\Psi(\bf{k}_i)\rangle\right]\\
&\approx \frac{1}{2}g_{11}(\bf{k})\varepsilon_1^2 +\frac{1}{2}g_{22}(\bf{k}+\varepsilon_1 \hat{\bf{x}}_1)\varepsilon_2^2 +\frac{1}{2}g_{11}(\bf{k}+\varepsilon_2 \hat{\bf{x}}_2)\varepsilon_1^2 \\
&+\frac{1}{2}g_{22}(\bf{k})\varepsilon_2^2\\
&\approx g_{11}(\bf{k})\varepsilon_1^2 +g_{22}(\bf{k})\varepsilon_2^2.
\end{align*}
Hence
\begin{align*}
&\prod_{i=1}^{4}\langle \Psi(\bf{k}_{i+1})|\Psi(\bf{k}_i)\rangle\\
&\approx\exp\left[-\left(g_{11}(\bf{k})\varepsilon_1^2 +g_{22}(\bf{k})\varepsilon_2^2\right)-F_{12}(\bf{k})\varepsilon_1\varepsilon_2\right].
\end{align*}

\section{No-go theorem: impossibility of having nontrivial flat K\"{a}hler bands for finite total number of bands}
\label{sec: no-go theorem}
Suppose we are given a K\"ahler band described by a holomorphic map $f:\BZ^2\to \mathbb{C}P^{n-1}; z\mapsto [Z_1(z):\dots:Z_{n}(z)]$, with $n$ the number of bands, where the map is holomorphic with respect to a flat complex structure determined by $\tau=\tau_1+i\tau_2 \in\mathcal{H}$ and complex coordinate $z=k_1+\tau k_2$. Note that the orthogonal projector $P(\bf{k})$ describing the band associated to $f$ is simply
\begin{align*}
P(\bf{k})=\sum_{i,j=1}^{n}\frac{Z_{i}(z)\overline{Z_j(z)}}{\sum_{k=1}^{n}|Z_k(z)|^2}\ket{i}\bra{j},
\end{align*}
where $\ket{i}$, $i=1,\dots,n$, is the canonical basis of $\mathbb{C}^{n}$ describing the internal degrees of freedom.

We want to show that $f^*\omega_{FS}$ cannot be uniform in the Brillouin zone. We want to compare the compatible tripe $(f^*\omega_{FS},j_{\tau},f^*g_{FS})$ and the flat anisotropic one given by 
\begin{align*}
 \omega&=\sqrt{\det (g)}dk_1\wedge dk_2=\frac{\mathcal{C}}{4\pi}dk_1\wedge dk_2=\frac{\mathcal{C}}{4\pi}\frac{1}{\bar{\tau}-\tau}dz\wedge d\overline{z},\\
 J &=\left[\begin{array}{cc}
-\frac{\tau_1}{\tau_2} & -\frac{|\tau|^2}{\tau_2}\\
\frac{1}{\tau_2} & \frac{\tau_1}{\tau_2}
\end{array}\right], \\
g &=\frac{\mathcal{C}}{4\pi}\left(\frac{-2i}{\overline{\tau}-\tau}\right)|dz|^2,
\end{align*}
where
\begin{align*}
\mathcal{C}=\int_{\BZ^2}\frac{f^*\omega_{FS}}{\pi}>0.
\end{align*}
We have $\mathcal{C}>0$ because the map $f$ is K\"{a}hler by assumption. Observe that $\omega=(-i/2)\partial \overline{\partial}\log h$, for $h=e^{-\frac{\mathcal{C}}{2\pi}\tau_2 k_2^2}$.
For the comparison we take the difference,
\begin{align}
\omega -f^*\omega_{FS}=-\frac{i}{2}\partial \overline{\partial} \log F,
\label{eq: comparison of symplectic forms}
\end{align}
where 
\begin{align}
F=|C(z)|^2\sum_{j=1}^{n}e^{-\frac{\mathcal{C}}{2\pi}\tau_2 k_2^2}|Z_{j}(z)|^2,
\end{align}
for some $C(z)$ holomorphic (not periodic function), defines a smooth function on $\BZ^2$. From the point of view of the formula of Eq.~\eqref{eq: comparison of symplectic forms}, the $C(z)$ may seem arbitrary, however, by the $\partial\overline{\partial}-$lemma (Corollary~3.2.10 of Ref.~\cite{huy:05}), it can be chosen so that $F$ is indeed a globally defined smooth function, because $\omega$ and $f^*\omega_{FS}$ lie in the same de Rham cohomology class (because they integrate to the same value $\pi \mathcal{C}$). As pointed out in Ref.~\cite{lee:cla:tho:17}, the uniformity of Berry curvature amounts to finding a map such that the K\"{a}hler potential for $f^*\omega_{FS}$ is Laplacian free. They also point out in the Appendix that, in their prescription, they would need an infinite number of parameters. Here we will go a bit further, and show that for finite total number of bands $n$ it is impossible to have a flat K\"{a}hler band. Since if the symplectic form is flat so will be the quantum metric by compatibility, it is enough to consider the flatness of the former. Note that the functions $b_j(z)=|C(z)|^2|Z_{j}(z)|^2$, $j=1,\dots,n$, must transform, under lattice translations, in a complementary way to $e^{-\frac{\mathcal{C}}{2\pi}\tau_2 k_2^2}$. They must also transform in a way which is independent of $j$, because for the map $f$ to be well defined, $Z_{j}(z+2\pi m+2\pi n\tau)/Z_{j}(z)$ must be independent of $j$, for all $m,n\in\mathbb{Z}$. Put differently, the quantities $C(z)Z_{j}(z)$, $j=1,\dots,n$, must transform as holomorphic sections of $L^{\otimes \mathcal{C}}$, where $L$ is the basic line bundle as defined in Sec.~\ref{sec: Explicit construction}. In particular, it means that $C(z)Z_{j}(z)$, $j=1,\dots,n$, can be written as linear combinations of a basis of $H^0(\BZ^2,L^{\otimes\mathcal{C}})$ whose dimension is, by Riemann-Roch, $\dim H^0(\BZ^2,L^{\otimes\mathcal{C}})=\mathcal{C}$. Making the replacements $p\leftrightarrow \mathcal{C}$ in the formulas of Sec.~\ref{sec: Explicit construction}, we can then write, using the basis of theta functions,
\begin{align*}
C(z)Z_{j}(z)&=\sum_{l=0}^{\mathcal{C}-1}A_{j}^{l}\vartheta\left[\begin{array}{c}
\frac{l}{\mathcal{C}}\\
0
\end{array}
\right]\left(\mathcal{C}\frac{z}{2\pi},\mathcal{C} \tau \right)\\
&=\sum_{l=0}^{\mathcal{C}-1}A_{j}^{l}a_{l}(z),
\end{align*}
where $A_{j}^{l}$ are complex numbers. Thus,
\begin{align}
F=\sum_{j=1}^{n}\sum_{l,l'=0}^{\mathcal{C}-1} e^{-\frac{\mathcal{C}}{2\pi}\tau_2 k_2^2}A_{j}^{l}\overline{A_{j}^{l'}}a_{l}(z)\overline{a_{l'}(z)}.
\label{eq: F}
\end{align}
We remark that the fact that there are $\mathcal{C}$ independent holomorphic sections ($\dim H^0(\BZ^2,L^{\otimes\mathcal{C}})=\mathcal{C}$) shows that for $n>\mathcal{C}$ the matrix $A=[A_{j}^{l}]_{1\leq j\leq n, 0\leq l\leq \mathcal{C}}$ is singular. This means that, at most, there are only $\mathcal{C}\times \mathcal{C}$ linearly independent coefficients. Thus, without loss of generality, we may assume that the sum on $j$ goes only up until $\mathcal{C}$.
The requirement for uniformity of $f^*\omega_{FS}$ is now seen to be equivalent to $F$ being a constant. We can even determine what that constant must be, because
\begin{align*}
\frac{1}{(2\pi)^2}\int_{BZ^2}F(z)dk_1\wedge dk_2 &= \sum_{j=1}^{n}\sum_{l,l'=0}^{\mathcal{C}-1}A_{j}^{l}\overline{A_{j}^{l'}}\langle s_{l},s_{l'}\rangle_{L^2}\\
&=\sum_{j=1}^{n}\sum_{l,l'=0}^{\mathcal{C}-1}A_{j}^{l}\overline{A_{j}^{l'}}\delta_{ll'}\sqrt{\frac{i}{\mathcal{C}(\tau-\bar{\tau})}}\\
&=\sum_{j=1}^{n}\sum_{l=0}^{\mathcal{C}-1}|A_{j}^{l}|^2\sqrt{\frac{i}{\mathcal{C}(\tau-\bar{\tau})}}.
\end{align*}

The quantity $a_{l}(z)\overline{a_{l'}(z)}$ can be written as
\begin{align*}
&a_{l}(z)\overline{a_{l'}(z)}\nonumber\\
&=\left(\sum_{m,n\in\mathbb{Z}} e^{i\frac{\pi \tau }{\mathcal{C}}\left(l+m\mathcal{C}\right)^2-\frac{i\pi \overline{\tau}}{\mathcal{C}}\left(l'+n\mathcal{C}\right)^2 +iz m- i\bar{z} n)\mathcal{C}}\right)e^{iz l-i zl'}
\end{align*}
The dependence $e^{iz l-i zl'}$, which gives a factor $e^{ik_1(l-l')}$, is the only $k_1-$dependence on the whole of $e^{-\frac{\mathcal{C}}{2\pi}\tau_2k_2^2}a_{l}(z)\overline{a_{l'}(z)}$ and, together with the fact that $a_{l}$ is $\mathcal{C}-$periodic, i.e., $a_{l}=a_{l+\mathcal{C}}$, and $\int e^{ik_1(l-l')}dk_2=2\pi \delta_{ll'}$, justifies why $\langle s_l,s_{l'}\rangle_{L^2}\propto \delta_{ll'}$. Furthermore, this same dependence implies that if $F$ were to be constant, the sum over $0\leq l,l'\leq \mathcal{C}-1$ must be restricted to the diagonal $l=l'$, otherwise there will be independent oscillatory terms in $k_1$ which will not allow for a constant $F$. Hence, we have reduced $F$ to be of the form
\begin{align}
F&=\sum_{j=1}^{n}\sum_{l=0}^{\mathcal{C}-1} e^{-\frac{\mathcal{C}}{2\pi}\tau_2 k_2^2}|A_{j}^{l}|^2 |a_{l}(z)|^2 \nonumber\\
&=\sum_{l=0}^{\mathcal{C}-1} e^{-\frac{\mathcal{C}}{2\pi}\tau_2 k_2^2}|A_{l}|^2 |a_{l}(z)|^2,
\label{eq: F reduced 1}
\end{align}
where we defined $|A_l|^2=\sum_{j=1}^{n}|A_{j}^{l}|^2$ for the sake of simplicity. We will now show that $F$ cannot be constant, by showing that the integral over $k_1$ defines a nonconstant function. Note
\begin{align*}
\int_{0}^{2\pi} \frac{dk_1}{2\pi} F(k_1,k_2)= \sum_{l=0}^{\mathcal{C}-1}|A_l|^2 \int_{0}^{2\pi} \frac{dk_1}{2\pi}    e^{-\frac{\mathcal{C}}{2\pi}\tau_2 k_2^2}|a_l(z)|^2.
\end{align*}
We then first  compute 
\begin{align*}
 &\int_{0}^{2\pi}\frac{dk_1}{2\pi} e^{-\frac{\mathcal{C}}{2\pi}\tau_2 k_2^2}|a_l(z)|^2\\
 &=\int_{0}^{2\pi}\frac{dk_1}{2\pi} e^{-\frac{\mathcal{C}}{2\pi}\tau_2 k_2^2}\\
 &\times \sum_{m,n\in\mathbb{Z}} e^{i\frac{\pi \tau }{\mathcal{C}}\left(l+m\mathcal{C}\right)^2-\frac{i\pi \overline{\tau}}{\mathcal{C}}\left(l+n\mathcal{C}\right)^2 +iz(l +m\mathcal{C})- i\bar{z} (l+ n\mathcal{C})}\\
 &=e^{-\frac{\mathcal{C}}{2\pi}\tau_2 k_2^2}\sum_{m\in\mathbb{Z}} e^{i\frac{\pi \tau }{\mathcal{C}}\left(l+m\mathcal{C}\right)^2-\frac{i\pi \overline{\tau}}{\mathcal{C}}\left(l+m\mathcal{C}\right)^2 +i(z-\bar{z})(l+m\mathcal{C})}\\
 &=e^{-\frac{\mathcal{C}}{2\pi}\tau_2 k_2^2}\sum_{m\in\mathbb{Z}} e^{-\frac{2\pi \tau_2 }{\mathcal{C}}\left(l+m\mathcal{C}\right)^2 -2\tau_2 k_2 (l+m\mathcal{C})}\\
 &=\sum_{m\in\mathbb{Z}}e^{-\frac{\mathcal{C} \tau_2}{2\pi }\left(k_2 +\frac{2\pi}{\mathcal{C}}\left(l+m\mathcal{C}\right)\right)^2}.
\end{align*}
Thus, we conclude that
\begin{align}
\int_{0}^{2\pi} \frac{dk_1}{2\pi} F(k_1,k_2)= \sum_{l=0}^{\mathcal{C}-1}|A_l|^2 \sum_{m\in\mathbb{Z}}e^{-\frac{\mathcal{C} \tau_2}{2\pi }\left(k_2 +\frac{2\pi}{\mathcal{C}}\left(l+m\mathcal{C}\right)\right)^2},
\end{align}
which is a sum of Gaussians centered at the points $\frac{2\pi}{\mathcal{C}}l+2\pi m$, with $l=0,\dots,\mathcal{C}-1$ and $m\in\mathbb{Z}$ (ensuring periodicity in $k_2$). Provided the $A_{l}$'s are nontrivial, which needs to happen for the map $f$ to be well-defined (there are further constraints on the $A_l$'s as for the map to be well-defined the $Z_j$'s cannot vanish or have poles simultaneously, but this is not relevant for the proof), this sum cannot be made constant in the case $\mathcal{C}$ is finite and the result is, thus, proved: there are no flat K\"{a}hler bands for finite total number of bands.

\bibliography{bib.bib}

\end{document}